\documentclass[useAMS]{mn2e}
\usepackage{graphicx}
\usepackage{txfonts}

\title[Dust in SDSS J1148+5251]{The origin of dust in high redshift QSOs: the case of SDSS J1148+5251}
\author[R. Valiante, R. Schneider, S. Salvadori, S. Bianchi]{Rosa Valiante$^{1,2}$\thanks{E-mail:
valiante@arcetri.astro.it}, Raffaella Schneider$^{3}$, Stefania Salvadori$^{4}$ and Simone Bianchi$^{2}$ \\
$^{1}$Dipartimento di Astronomia, Universita' di Firenze, Largo Enrico Fermi 2, 50125 Firenze, Italy\\
$^{2}$INAF - Osservatorio Astrofisico di Arcetri, Largo Enrico Fermi 5, 50125, Firenze, Italy\\
$^{3}$INAF - Osseravtorio Astronomico di Roma, via di Frascati 33, 00040, Monteporzio Catone, Italy\\
$^{4}$Kapteyn Astronomical Institute, Landlaven 12, 9747 AD Groningen, the Netherlands}
\begin{document}

\date{Accepted . Received }

\pagerange{\pageref{firstpage}--\pageref{lastpage}} 
\pubyear{2011}

\maketitle

\label{firstpage}

\begin{abstract}
We present a semi-analytical model for the formation and evolution of a high
redshift quasar (QSO). 
We reconstruct a set of hierarchical merger histories of a $10^{13}$ M$_\odot$ 
dark matter halo and model the evolution of the corresponding galaxy and of 
its central super massive black hole. The code \textsc{GAMETE/QSOdust} 
consistently follows: 
{\it (i)} the black hole assembly via both coalescence with other black holes 
and gas accretion; 
{\it (ii)} the build up and star formation history of the quasar host 
galaxy, driven by binary mergers and mass accretion; 
{\it (iii)} the evolution of gas, stars, metals in the interstellar medium (ISM), 
accounting for mass exchanges with the external medium (infall and outflow processes);
{\it (iv)} dust formation in Supernova (SN) ejecta and in the stellar atmosphere of
Asymptotic Giant Branch (AGB) stars, dust destruction by interstellar
shocks and grain growth in molecular clouds; 
{\it (v)} the AGN feedback which powers a galactic-scale wind, self-regulating the 
black hole growth and eventually halting star formation.\\
\noindent
We use this model to study the case of SDSS J1148+5251 observed at 
redshift 6.4.
We explore different star formation histories for the QSO 
host galaxy and find that Population III stars give a negligible 
contribution to the final metal and dust masses due to rapid 
enrichment of the ISM to metallicities $> Z_{\rm cr} = 
[10^{-6} - 10^{-4}] Z_{\odot}$ in progenitor galaxies at redshifts $>10$. 
If Population II/I stars form with a standard initial mass function (IMF) 
and with a characteristic stellar mass of $m_{ch}=0.35$ M$_\odot$,
a final stellar mass of $[1-5]\times 10^{11} M_{\odot}$ 
is required to reproduce the observed dust mass and gas metallicity of J1148. 
This is a factor 3 to 10 higher than the stellar mass inferred from observations 
and would shift the QSO closer or onto the stellar bulge - black hole relation
observed in the local Universe; alternatively, the observed chemical 
properties can be reconciled with the inferred stellar mass assuming 
that Population II/I stars form according to a top-heavy IMF with 
$m_{ch}=5$ M$_\odot$. We find that SNe dominate the early dust enrichment 
and that, depending on the shape of the star formation history and on the
stellar IMF, AGB stars contribute at redshift $z < 8 - 10$. 
Yet, a dust mass of $[2-6] \times 10^8$ M$_\odot$ estimated for J1148
cannot be reproduced considering only stellar sources and the final dust
mass is dominated by grain growth in molecular clouds. This conclusion is
independent of the stellar initial mass function and star formation history. 
\end{abstract}

\begin{keywords}
Galaxies: evolution, high-redshift, ISM; quasars: general; stars: AGB and 
post-AGB, supernovae: general, ISM: dust, extinction 
\end{keywords}

\section{Introduction}
The presence of large dust masses in high redshift quasars (QSOs) 
has been revealed by mm and sub-mm observations of samples of $5<z<6.4$ QSOs
in the Sloan Digital Sky Survey (SDSS) (Bertoldi et al. 2003; Priddey et al.
 2003; Robson et al. 2004; Beelen et al. 2006; Wang et al. 2008). 
The detection of dust thermal emission and the inferred far-IR (FIR) 
luminosities of these objects suggest masses of warm 
($T<100$K) dust of a few $10^{8}$M$_{\odot}$.
Observations of QSOs and galaxies at high redshift are strongly affected by 
the presence of dust and our ability to interpret the observed properties of 
high redshift sources depends on a detailed understanding of the star 
formation history and the history of metal and dust pollution in their 
interstellar medium (ISM). Dust properties, such as grains chemical composition and size 
distribution, are of primary importance in modeling the extinction curves in 
both the local Universe (Cardelli, Clayton \& Mathis 1989; Pei 1992; 
Weingartner \& Draine 2001) and at high redshifts (Maiolino et al. 2004; 
Stratta et al. 2007; Gallerani et al. 2010). 
Yet, the origin of dust at such early epochs is still unclear. 

Among the stellar sources, the cool and dense atmosphere of
Asymptotic Giant Branch (AGB) stars and the expanding ejecta of core collapse 
supernova (SN) offer the most viable sites of dust grains condensation. 
A SN origin has often been advocated as the only possible explanation for the 
large amount of dust observed in high redshift QSOs since it is usually 
assumed that low and intermediate-mass stars ($m_{\ast}<8$M$_{\odot}$) must 
evolve on a timescale comparable to the age of Universe at redshift $z\geq 6$ 
(about 1~Gyr, Morgan \& Edmunds 2003; Marchenko 2006; Dwek, Galliano \& Jones 
2007) before they reach the dust producing stages. 
This scenario has been tested through observations of the reddened 
quasar SDSSJ1048+46 at $z = 6.2$ (Maiolino et al. 2004) and of the Spectral 
Energy Distribution (SED) of the $z =6.29$ GRB 050904 (Stratta et al. 2007) and 
the $ z \sim 5$ GRB 071025 (Perley et al. 2010). In these sources, 
the inferred dust extinction curve is different with respect 
to any of the extinction curves observed at low $z$, 
and it shows a very good agreement with the extinction curve predicted for 
dust formed in SN ejecta. This is an indication that the properties of dust 
may evolve beyond $z \ge 5 - 6$. In a recent study, Gallerani et al. (2010) find that the 
extinction curves of a sample of QSOs with $4\leq z\leq 6.4$ deviate from the 
Small Magellanic Cloud (SMC) 
extinction curve (the one which better describe the $z<2$ quasar reddening),
suggesting that production mechanisms and/or dust processing
into the ISM must be different at high redshift. 

On the theoretical side, the evolution of dust in the early Universe is still 
poorly understood. 
Some of the models developed so far are still consistent with the scenario 
where most of the dust in QSO at $z \sim 6$ is produced by SNe. However these 
estimates either neglect dust destruction by interstellar shocks (Maiolino et 
al. 2006; Li et al. 2008) or make extreme assumptions on SN dust condensation 
factors (Dwek, Galliano \& Jones 2007).
In addition, it has been shown that only $[2-20]$\% 
of the newly condensed dust survives the passage of the SN reverse shock and 
grains are reduced both in number and in size, depending on
the surrounding ISM density as well as on dust composition (see Bianchi \& Schneider 
2007; Nozawa et al. 2007; Bianchi et al. 2009).

In a previous paper (Valiante et al. 2009; hereafter V09) we showed that 
AGB stars can provide a non negligible contribution to dust formation
at high $z$, being able to produce dust already $[30-40]$~Myr after 
the onset of the star formation, which is about the mean lifetime of a 
$\le 8~$M$_{\odot}$ star (see e.g. Padovani \& Matteucci 1993; Raiteri et al. 1996). 
In particular, for a standard stellar initial mass function (IMF), 
the characteristic timescale at which AGB stars dominate dust production
ranges between 150 and 500 Myr, depending on the assumed
star formation history and on the stellar initial metallicity. 
Hence, we concluded that these stellar dust sources must be taken into account
when modeling dust evolution at high redshift.
In V09 we applied a simple chemical evolution model with dust to the host
galaxy of the $z = 6.4$ SDSS J1148+5251 QSO (hereafter J1148) using as an 
input to the model the star formation history obtained in a numerical 
simulation by Li et al. (2007); this simulation predicts that a large stellar 
bulge is already formed at $z = 6.4$, in agreement with the local SMBH-stellar 
mass relation, through a series of strong bursts triggered by galaxy mergers 
in the hierarchical evolution of the system. We found that the total mass of 
dust predicted at $z = 6.4$ is within the range of values inferred by 
observations of SDSS J1148, with a substantial contribution (80\%) of AGB-dust.
We concluded that stellar sources can account for the
huge dust mass observed at these early cosmic epochs, even when dust 
destruction by SN shocks is taken into account, at the price of having a total 
stellar mass of $\sim 10^{12} \rm{M}_{\odot}$, which can hardly be reconciled 
with the dynamical mass of $\sim 5 \times 10^{10} \rm{M}_{\odot}$ inferred from 
CO observations within 2.5 kpc from the QSO center (Walter et al. 2004). 

The above critical point has been recently emphasized by Michalowski et al. 
(2010) who showed that using stellar masses approximated as the difference 
between the dynamical and the H$_2$ gas masses for $5 < z < 6.5$ QSOs, stellar 
sources of dust are not efficient enough to account for the dust mass inferred 
in these systems. Therefore, they suggest additional non-stellar dust formation
 mechanism; in particular, significant dust grain growth in the interstellar 
medium of these QSOs. 
Indeed, dust grains formed in stellar outflows can serve as growth centers for 
accretion of icy mantels in dense Molecular Clouds (MC, see e.g. Draine 1990).
It is already well known that MC-grown dust is the dominant component in the 
Milky Way (see e.g. Zhukovska et al. 2008). Given the large masses of molecular
gas detected in CO for most of these systems, Draine (2009) suggested that the 
bulk of their dust mass is the result of grain growth and that SN are required 
to produce the metals that compose the grains and the ''seed" grain surface 
area on which more material can grow in MCs.   

With the present work we investigate the nature of dust in
high-$z$ QSOs host galaxies focusing in particular on J1148 at $z=6.4$.
To this aim, we have improved the semi-analytic code \textsc{GAlaxy MErger 
Tree and Evolution (GAMETE)}, developed by Salvadori, Schneider \& Ferrara 
(2007, hereafter SSF07) to 
model the hierarchical assembly and merger history of a QSO host
galaxy at $z=6.4$, together with its central super massive
black hole including feedback from the active galactic nucleus (AGN) in the form 
of a galactic-scale wind. The chemical evolution network has also been 
modified to describe the evolution of dust taking into account the specific
stellar lifetimes, and the evolution of dust in the ISM through destruction
by interstellar shocks and dust-growth in MCs. 
This code, which we will refer to as \textsc{GAMETE/QSOdust}, will enable us
to overcome the main limitations of V09 analysis, namely: {\it (i)} we will
be able to model different possible star formation histories for J1148 
discussing their impact on the chemical properties of the QSO host and, in 
particular, on the final dust mass, and {\it (ii)} we will be able to explore the 
relative importance of stellar sources of dust (SN and AGB) and of grain growth 
in MCs. At the same time, the model can be constrained using the properties 
directly observed or inferred from observations of J1148, such as the mass of 
the central black hole, the total mass of gas and stars, the mass in metals and dust. 

The paper is organized as follows:
in section 2 we will introduce the main physical properties of the QSO J1148 and
its host galaxy discussing the observational uncertainties.  
In section 3 we will briefly summarize the basic features of the original 
GAMETE code and present the parameters adopted to simulate the hierarchical 
assembly of a QSO host halo at $z = 6.4$. 
In section 4 we will discuss the new features of the code which
allow to follow the formation and evolution of the central black hole and its 
effects on the parent host galaxy through AGN-driven feedback. 
In section 5 we describe the chemical evolution network, including the 
evolution of dust in the ISM. In section 6 we will present 
the results and in section 7 we will discuss their implications. 
Finally, in section 8 we will draw the main conclusions of this work.

In what follows we adopt a Lambda Cold Dark Matter ($\Lambda$CDM) 
cosmology  with $\Omega_m = 0.24$, $\Omega_\Lambda = 0.76$, $\Omega_b = 0.04$, 
and $H_0 = 73$~km/s/Mpc. 
The age of the Universe at a redshift $z = 6.4$ is 900 Myr. 

\section{J1148: a QSO at redshift 6.4}
\label{sec:obs}

In this section we briefly summarize the physical properties of this 
extremely luminous QSO and of its host galaxy which have been derived 
observationally. The full list of properties, and observational uncertainties,
is provided at the end of this section.

\subsection{Dark matter host halo mass and AGN properties}
The mass of the dark matter halo where J1148 is hosted, $M_{\rm h}$, is usually 
estimated by requiring that the number density in halos with masses 
$\ge M_{\rm h}$ at $z \sim 6$ matches the observed space density of QSOs at the 
same redshift (Fan et al. 2004). Depending on the cosmological model, this 
generally yields to $M_{\rm h} = [10^{12} - 10^{13}] M_{\odot}$. 
The space density of observed QSOs with average redshift $<z> = 6$ and 
luminosity M$_{1640} < -26.7$ is $(6.4 \pm 2.4) \times 10^{-10}$~Mpc$^{-3}$ 
(Fan et al. 2004). 
For our adopted cosmological model, this would require a minimum halo mass of 
$\sim 3 \times 10^{12} M_{\odot}$. 
To compare with previous analyses (Volonteri \& Rees 2006; Valiante et al. 
2009) in this work we generate Monte Carlo realizations of the merger 
histories adopting a value of $M_{\rm h} = 10^{13} M_{\odot}$. 
The implications of a different halo mass and redshift will be addressed in a 
forthcoming paper, where a larger sample of QSOs will be investigated. 

J1148 is a very luminous QSO with absolute AB magnitude of the continuum in the 
rest-frame at 1450 \AA~of $M_{1450} = -27.82$ (Fan et al. 2003). Assuming a 
power law $f_{\nu} \propto \nu^{\alpha}$  with $\alpha = - 0.5$ the corresponding 
B-band magnitude\footnote{Schmidt, Schneider \& Gunn (1995) find 
$M_{\rm B} = M_{1450} +2.5 \alpha \rm{Log} (4400 \AA/1450\AA) +0.12$ 
where the factor 0.12 comes from the zero point difference between the AB and 
the Vega-based magnitude system for QSO-like spectra.} 
and luminosities are $M_{\rm B} = -28.3$ and $L_{\rm B} = 5.2 \times 10^{12} 
L_{\odot}$ where the effective wavelength of the B-band is $4400 \AA$. 

The energy requirement to power such luminosity suggests the presence of a 
black hole mass of at least several billion solar masses. A virial estimate of 
the black hole mass of $3\times 10^9\rm{M_\odot}$ was obtained by Willott et al.
 (2003) using the Mg$_{\rm II}$ $\lambda$2800 line-width and the scaling 
relations derived by McLure \& Jarvis (2002). Using the same scaling relations 
but also the virial estimator provided by Vestergaard (2002), Barth et al. 
(2003) consistently found a BH mass in the range $(2-6) \times 10^{9} M_{\odot}$ 
from both the widths of the C$_{\rm IV}$ $\lambda$1549 and Mg$_{\rm II}$ emission 
lines. Beside the factor 2.5-3 estimated uncertainty in the adopted scaling 
relations, this method assumes that the dynamic of the line-emitting gas is 
dominated by gravitational forces. Note, however, that for 
objects radiating close to the Eddington limit, the effect of radiation 
pressure cannot be neglected and may lead to a non negligible upward 
correction of the estimated BH masses (Marconi et al. 2008; Lamastra et al. 
2010).   

\subsection{Host galaxy properties}\label{sec:hostprop}
High-resolution VLA observations of J1148 have enabled to resolve CO (3-2)
line emission both spatially and in terms of velocity (Walter et al. 2004).
The molecular gas is extended to a radius of $\sim 2.5$ kpc from the center
and a total mass of $M_{\rm H_2} = 1.6 \times 10^{10} M_\odot$ has been 
derived, assuming a CO intensity-to-gas mass conversion factor appropriate
for ultra luminous infrared galaxies (ULIRGs, Walter et al. 2004).
Assuming that the gas is gravitationally bound and forms an inclined disk
with inclination angle $i$, Walter et al. (2004) infer a dynamical mass of 
$M_{\rm dyn} sin^2i \sim 4.5 \times 10^{10} \rm{M_\odot}$.
The molecular disk inclination angle is the main source of uncertainty in the 
dynamical mass estimates (Wang et al. 2010). Walter et al (2004) correct for 
an inclination of $65^\circ$, on the basis of the broad CO FWHM detected, 
obtaining $M_{\rm dyn} \sim 5.5 \times 10^{10} M_{\odot}$ with 50\% uncertainty.   

The derivation of the stellar bulge mass is extremely uncertain. Indeed this 
can be either estimated by extrapolating the local $M_{\rm BH}-M_{\ast}$ relation 
(Marconi \& Hunt 2003; Haring \& Rix 2004) at high redshift, 
$M_{\rm BH}/0.0014 \sim 2.14 \times 10^{12} M_{\odot}$ (if a BH mass of 
$3 \times 10^9 M_{\odot}$ is considered), or by subtracting from the 
observed dynamical mass the H$_2$ gas mass, $M_\ast \sim M_{\rm dyn} - M_{\rm H_2}
\sim 3.9 \times 10^{10}$ M$\odot$. It is evident that these estimates are 
unconsistent, since they differ by almost two orders of magnitude.
Such a discrepancy persists, though alleviated, even if we assume that the stellar
bulge mass of the host galaxy of J1148 was 10 times the scale length of the
molecular gas emission. Assuming a density profile of
$\rho \propto r^{-2}$, the stellar mass would be  
$M_{\ast} \sim 3.9 \times 10^{11} \rm{M_\odot}$ (Walter et al. 2004). 

Further constraints on the evolution of the QSO host galaxy 
come from observations of the gas phase metallicity and elemental abundances 
in both the broad and narrow line regions (BLRs, NLRs). 
BLRs metallicity tracers are emission lines ratios,
such as Fe\textsc{ii}/Mg\textsc{ii} (e.g. Barth et al. 2003), 
N\textsc{v}/C\textsc{iv} (e.g. Pentericci et al. 2002) and 
(Si\textsc{iv}+O\textsc{iv})/C\textsc{iv} (Nagao et al. 2006; Juarez et al.
2009), together with other metal lines like C\textsc{ii} (e.g. Maiolino et al. 
2005), and O\textsc{i} (Becker et al. 2006). 
The observed ratios are similar to those observed in low-$z$ QSOs, 
indicating BLRs metallicities which are several times solar, $\sim 7~\rm{Z_\odot}$ (Nagao et al. 2006, Juarez et al. 2009) and  
suggesting that the ISM of the QSO host was significantly enriched by heavy 
elements.  
However, the BLRs are small nuclear regions, of less than a few pc, containing a 
total mass of $\sim 10^4 M_{\odot}$; thus, we consider such a high metallicity
as an upper limit.
Conversely, NLRs have larger masses and a size which is roughly comparable 
to the size of the host galaxy and represent a good tracer of the chemical 
properties on galactic scales, since their metal content reflects the past 
star formation history of the galaxy  (Matsuoka et al. 2009 and references therein). 
To date however, only measurements of the NLRs metallicity up to redshift 4 are 
available. Focusing on the C\textsc{iv}/He\textsc{ii} and 
C\textsc{iii}/C\textsc{iv} flux ratios, Matsuoka et al. (2009) point out that 
even NLRs show no significant metallicity evolution with
redshift, up to $z\sim 4$. For the highest redshift sample these authors find a
 NLR metallicity of about $1.32^{+1.57}_{-1.10} \rm{Z_\odot}$. We adopt this 
value as a lower limit for the ISM metallicity of J1148.  

\subsection{Dust mass, FIR luminosity and SFR}\label{sec:dustmass}
The dust present in the host galaxy is probably characterized by 
multi-temperature components: hot ($\sim 100$ K), warm ($30-80$ K) and  
cold ($10-20$ K). 
The hot dust component, directly heated by the AGN activity, has been observed 
for several $z\sim 6$ QSOs with Spitzer (Charmandaris et al. 2004; Jiang et 
al. 2010) and represents the least contribution to the total mass of dust.  
Warm dust, associated to starburst regions, dominates the emission in the 
rest-frame far-IR. For J1148, it has been observed in 4 bands:
at 1.2 mm (Bertoldi et al. 2003), 850$\mu$m, 450$\mu$m (Robson et al. 2004), 
and 350$\mu$m (Beelen et al. 2006). 
The cold dust, finally, may represent the largest component. Its emission, 
however, would be hidden by the warmer component even if its mass were 
$1.5-3$ times larger (Dunne et al. 2000).  

In the rest-frame far-IR the emission is optically thin, $\tau_{\rm d}(\lambda) << 1$, and the
dust mass can be estimated from the flux observed in a given band $S_{\nu_0}$ as,
\begin{equation}
M_{\rm dust} = \frac{S_{\nu_0} \, d_{\rm L}^2(z)}{(1+z) \, \kappa_{\rm d} (\nu) \, B(\nu, T_{\rm d})},
\label{eq:mdust}
\end{equation}
\noindent
where $\kappa_{\rm d}(\nu)$ is the opacity coefficient per unit dust mass, $B(\nu, T_{\rm d})$
is the Planck function for a dust temperature $T_{\rm d}$, and $d_{\rm L}$ is the luminosity
distance to the source. In the Rayleigh-Jeans part of the spectrum, dust radiates as a 'grey-body'
with $\kappa_{\rm d}(\nu) = \kappa_0 (\nu/\nu_0)^\beta$. Integrating over the spectrum one obtains
the corresponding far-IR luminosity,
\[
L_{\rm FIR} = 4 \pi M_{\rm dust} \int \, \kappa_{\rm d} (\nu) \, B(\nu, T_{\rm d}) \, d\nu.
\]
\noindent
Although in principle with four photometric
points it could be possible to simultaneously fit $M_{\rm dust}$, $\beta$, and $T_{\rm d}$, 
at the redshift of J1148 the observations do not fully sample the Rayleigh-Jeans part of the spectrum,
preventing a determination of $\beta$. Therefore, the estimated mass of dust and FIR luminosities 
for J1148 quoted in the literature have been derived under different assumptions for $\beta$ and
$\kappa_0$. The resulting dust masses, FIR luminosities and dust temperatures are listed in Table
\ref{table:dustmass} and have been calculated through a $\chi^2$ fit using all the observed points
in the range $350$~$\mu$m - $1200$~mm. 
Taking into account the intrinsic error on the estimated fluxes, the 
typical error associated to a single fit is $25\%$. 
As it can be inferred from the Table, we have considered, in addition to dust absorption coefficients
already adopted in previous works (cases a-c), a fit to the Weingartner \& Draine (2001) model for the
SMC in the relevant frequency range (case d) as well as the dust optical properties
expected for SN dust following the model by Bianchi \& Schneider (2007, case e).

\begin{table*}
\begin{center}
\caption{Dust temperature, dust mass, and FIR luminosity resulting from the $\chi^2$-fit of the
four observed photometric points of J1148 (see text). Each lines refer to the values
inferred from the fit taking the dust absorption coefficient as specified by the
corresponding $\kappa_0$ and $\beta$. These values have been chosen following quoted
values in the literature and assuming specific dust grain models. In particular:  
a = Bertoldi et al. (2003); b = Robson et al. (2004); c = Beelen et al. (2006); 
d = fit to the Weingartner \& Draine (2001) model for the SMC in the
spectral range [40-200]$\mu$m; e = SN dust model by Bianchi \& Schneider (2007).}\label{table:dustmass}
\begin{tabular}{l|l|l|l|l|l|l}\hline
Ref. & $\kappa_0$ [cm$^2$/gr]  & $\lambda_0$ [$\mu$m] & $\beta$ &  $T_{\rm d}$ [K] & $M_{\rm dust}$ [M$_{\odot}$]& $L_{\rm FIR}$ [$L_{\odot}$] \\ \hline
a  & 7.5 & 230 & 1.5 & 58 & $3.16 \times 10^{8}$ & $2.32 \times 10^{13}$ \\
b & 30 & 125 & 2.0 & 49 & $2.91 \times 10^{8}$ & $2.09 \times 10^{13}$ \\
c & 0.4 & 1200 & 1.6 & 56 & $4.29 \times 10^{8}$ & $2.27 \times 10^{13}$ \\
d  & 34.7 & 100 & 2.2 & 47 & $4.78 \times 10^8$ & $2.02 \times 10^{13}$ \\
e  & 40 & 100 & 1.4 & 60 & $1.86 \times 10^8$ & $2.38 \times 10^{13}$ \\
\hline
\end{tabular}
\end{center}
\end{table*}
  
The resulting dust masses and temperatures are very sensitive to the adopted dust 
absorption coefficient and spectral index with values in the range 
$[1.86 - 4.78] \times 10^8 M_{\odot}$. 
In what follows we will assume an average dust mass of 
$3.4 \times 10^8 M_{\odot}$ with errorbars computed assuming the minimum and 
maximum values quoted in the Table.
It is important to note that a mixture of SMC and SN extinction curves (30\% 
SMC and 70\% SN) is found to best fit the extinction properties of a sample of 
QSOs with $3.9 \le z \le 6.4$, including J1148 (Gallerani et al. 2010).   

From Table \ref{table:dustmass} we also infer an average FIR luminosity of $L_{\rm FIR} = 2.2 \times 10^{13} L_{\odot}$ with 15\% uncertainty associated to the fit. Assuming that the dominant dust heating mechanism is radiation from young stars, it is possible to estimate
the star formation rate (SFR) using the relationship between FIR luminosity and SFR derived by 
Kennicutt (1998)\footnote{The conversion factor $5.8\times 10^9$ has been derived assuming a mean luminosity for a 10-100 Myr 
burst, stars with solar abundances and a Salpeter IMF (see the original Kennicutt 1998 paper for details).},
\[
\frac{\rm SFR}{1 M_\odot {\rm yr}^{-1}} = \frac{L_{\rm FIR}}{5.8\times 10^9 L_{\odot}},
\]
\noindent
which implies a SFR $\sim (3.8\pm 0.57) \times 10^3 M_{\odot}$/yr (Bertoldi et al. 2003). 
Such a high $\rmn{SFR}$ is also supported by the first detection of the carbon 
[CII] line at 158 $\rm{\mu}$m (Maiolino et al. 2005), where a SFR $\sim 3 \times 10^3 M_{\odot}$/yr
is estimated combining the above conversion with the [CII] line - FIR luminosity ratio obtained
from PDR models.  

It is clear that if the AGN contributes to dust heating, the corresponding 
$\rmn{SFR}$ would be much lower, consistent with the value that would be 
inferred from the observed gas mass by a simple application of the Schmidt-Kennicutt
law. In fact, using the relation proposed by Daddi et al. (2010) to fit
local ULIRGs and submillimetre galaxies (SMGs) and QSOs, 
\[
{\rm Log} \Sigma_{\rm SFR} /[M_{\odot} {\rm yr}^{-1} {\rm kpc}^{-2}] = 1.42 \times {\rm Log} \Sigma_{\rm gas}/[M_{\odot}{\rm pc}^{-2}] - 2.93 
\]
\noindent
and assuming that the SFR is confined in the same central 2.5 kpc where 
molecular gas emission has been detected, we find a SFR $\sim 180 M_{\odot}$/yr 
consistent with previous findings by Dwek et al. (2007) and Li et al. (2007).\\ 
\newline
\noindent
Summarizing, the observed/inferred properties of J1148 used as reference values
in this work are:
\begin{itemize}
  \item A DM halo mass of $M_h=10^{13}$ M$_\odot$, which is assumed to host the
    $z=6.4$ QSO J1148.
  \item A SMBH mass of $M_{\rm BH}\sim 3^{+3.0}_{-1.0}\times 10^9$ M$_\odot$ 
    inferred from the Mg$_{\rm II}$ $\lambda$2800 and C$_{\rm IV}$ $\lambda$1549 
    line-widths (Willot et al. 2003; Barth et al. 2003).
  \item A dynamical and molecular gas masses of $M_{\rm dyn}=(5.5\pm 2.75)\times 
    10^{10}$ M$_\odot$ and $M_{\rm H_2}=1.6\times 10^{10}$ M$_\odot$, derived from 
    observations of the CO(3-2) line emission and assuming an inclination 
    angle $i=65^\circ$ (Walter et al. 2003).
  \item A final stellar mass ranging from $M_\ast = (3.9\pm 2.75)\times 
      10^{10}$ M$_\odot$ (computed as the difference between the dynamical and 
      molecular gas masses given above) up to $M_\ast \sim 10^{12}$ M$_\odot$ 
      which is the one inferred if the local $M_{\rm BH}-M_\ast$ relation were to 
      hold even at redhift 6.4.
  \item An ISM metallicity of $Z=1.32^{1.57}_{1.10}$ Z$_\odot$,
    estimated from observations of NLRs in high redshift QSOs (Matsuoka et al.
    2009). We refer to this value as a lower limit for the metallicity of J1148.
  \item A total mass of dust of $M_{\rm dust}=3.4^{+1.38}_{-1.54}\times 10^8$ M$_\odot$,
    computed using eq. (\ref{eq:mdust}) and the parameters given in Table 
    \ref{table:dustmass}.
  \item An upper limit to the SFR, $\sim (3.8 \pm 0.57)\times 10^3$ 
    M$\odot$/yr, obtained assuming that dust heating is entirely due to radiation
    from young stars.
  
\end{itemize}

\section{Model description}
\label{sec:model}

Here we describe our Monte Carlo semi-analytical model, 
\textsc{GAMETE/QSOdust}, which follows the formation and evolution of the 
quasar SMBH and of its host galaxy in the hierarchical scenario of structure 
formation. In this framework, a galaxy form through a series of merging 
episodes of lower mass fragments, called {\it progenitors}.
The code is divided into two main blocks: the first block runs backward in 
time and it is needed to reconstruct the merger tree history of the assumed 
$10^{13}\rm{M_\odot}$ DM halo at $z=6.4$ (see section \ref{sec:obs}); the second 
block runs forward in time and allows us to follow the evolution of the 
baryonic component and the growth of the nuclear black hole. 
In this way, it is possible to simulate several hierarchical merger histories 
of the host DM halo and then follow the build up of the quasar and the 
evolution of the host galaxy properties along these various formation paths.
Each formation path will result in a peculiar redshift evolution for 
the host galaxy properties. However, in order to draw some general 
conclusions, in what follows we will show the results obtained averaging over 
all the simulated hierarchical merger histories.

In the following sections we will briefly describe the Monte Carlo algorithm
used to reconstruct the hierarchical merger tree of the $10^{13}$ M$_\odot$ DM 
halo hosting the QSO at redshift $z=6.4$, (see SSF07 for a detailed description
of the algorithm) and the semi-analytical model which follows the evolution of 
gas, stars and metals along the hierarchical build-up of the galaxy (see 
Salvadori, Ferrara \& Schneider 2008, herafter SFS08, for more details). 
Then, we will introduce in details the new features 
of the \textsc{GAMETE/QSOdust} version implemented to follow the evolution of 
the central SMBH and of the dust mass. 
\subsection{The Hierarchical Merger History}\label{sec:mergertree}
The possible hierarchical merger histories of a $M_h=10^{13}$ M$_\odot$ DM halo 
at redshift $z=6.4$ are reconstructed using a {\it binary} Monte Carlo 
algorithm with {\it mass accretion} (e.g. Cole et al. 2000; Volonteri, Haardt 
\& Madau 2003), based on the Extended Press-Schechter (EPS) theory (e.g. Lacey 
\& Cole, 1993).

At each increasing redshift, a DM halo can either loose part of its mass or loose 
mass and fragment into two progenitors with random masses in the range 
$M_{\rm res}<M<M_h/2$, where $M_{\rm res}$ is a threshold mass representing 
the model resolution limit:
halos with $M<M_{\rm res}$, are not considered as progenitors but account for
the surrounding medium in which other halos are embedded and from which 
they accrete mass and virialize.

\begin{figure}
\includegraphics [width=8.0cm]{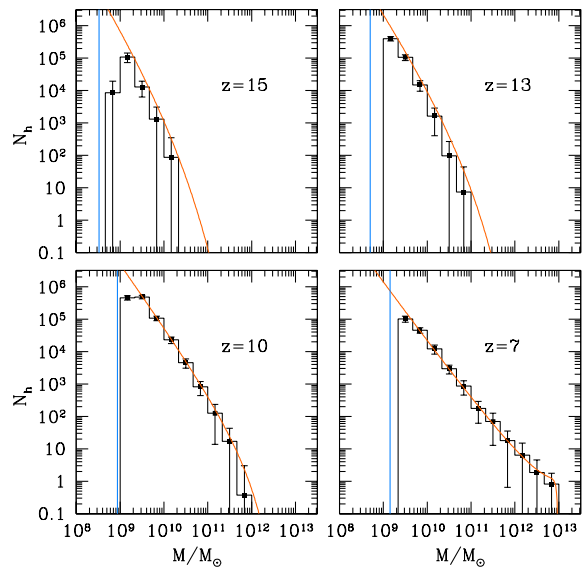}
\caption{Number of progenitors of a QSO host halo of mass $10^{13} 
\rm{M_\odot}$ at $z=6.4$ as a function of mass at four different redshifts:
$z = 15, 13, 10$ and $7$ (from top left to bottom right). In each panel, 
histograms represent averages over 50 realizations of the merger tree 
and the errorbars indicate Poissonian error on 
the counts in each mass bin; solid lines show the predictions of EPS theory 
and the vertical lines indicate the values of the resolution mass at the 
corresponding redshift.} 
\label{fig:epsreal} 
\end{figure}

The resolution mass, $M_{\rm res}$, along with the redshift interval chosen to
sample the merger history, represent the
free parameters of the model and are constrained in order to 
{\it (i)} prevent multiple fragmentations, 
{\it (ii)} resolve all the progenitors with masses above a given physical 
threshold, 
{\it (iii)} reproduce the EPS predictions,
{\it (iv)} control the computational cost.
Given the above requirements, we assume that $M_{\rm res}(z) = 10 M(T_{\rm vir}=10^4\rm{K},z)$, where $M(T_{\rm vir}=10^4 \rm{K},z)$ is the mass of a DM halo with a virial equilibrium temperature of $10^4\rm{K}$ at redshift $z$.
The merger tree is computed using 5000 redshift intervals logaritmically 
spaced in the expansion factor between $z=6.4$ and $z=37$. 
In Fig.~\ref{fig:epsreal} we compare the mass function of the progenitor halos 
averaged over 50 possible hierarchical merger histories with the EPS 
predictions. We can see that the agreement is extremely good at all
redshifts ($z = 15, 13, 10, 7$). 

We then use the produced merger trees as an input to 
reconstruct the formation and evolution of the quasar J1148 
and its host galaxy through cosmic time.     

\subsection{Evolution along a merger tree}\label{sec:model_summary}
In what follows we present a brief summary of the basic
prescriptions proposed by SFS08 which are unchanged in the present version of
the code. 
Note that in order to significantly reduce the high computational cost caused 
by the high number of redshift intervals, we are now considering as  
{\it star-forming} objects all DM halos with masses equal or greater than 
$M_{\rm sf}(z) = 10 M_{\rm res}(z)$ (see Fig.~\ref{fig:res}).

In any star-forming halo, the star formation rate is taken to be proportional
to the mass of cold gas,
\begin{equation}
{\rm SFR} = f_{\ast}(z) M_{\rm gas},
\label{eq:sfr}
\end{equation}
\noindent
whose gradual accretion is regulated by an infall rate, $dM_{\rm inf}/dt$ (see
SFS08 for the complete expression).
The parameter $f_{\ast}(z)$ is the redshift-dependent global efficiency of star 
formation per unit time. 
We will discuss this efficiency in detail in the next section.

In this model, Population III (Pop III) stars with an average 
mass of $200 M_\odot$ are assumed to form if the gas metallicity is lower than 
a critical value $Z_{\rm cr} = 10^{-4} Z_{\odot}$. 
Otherwise, if the gas metallicity exceeds the critical limit, low-mass 
(Pop II/I) stars are assumed to form according to a Larson 
initial mass function (IMF), which follows a Salpeter-like power law at the 
upper end but flattens below a characteristic stellar mass:
\begin{equation}
\phi(m) \propto m^{-(\alpha + 1)} e^{-m_{ch}/m},
\label{imf}
\end{equation}
\noindent
where $\alpha=1.35$, $m_{ch}=0.35 M_{\odot}$ and we normalize the integral of 
$m \phi(m)$ in the mass range $(0.1-100)$ M$_{\odot}$ to unity.
The evolution of very massive stars is rapid (they reach the end of their
main sequence phase in 3-5 Myr) and ends in a violent explosion as pair 
instability supernovae (PISN) which leave no remnants behind and significantly 
contribute to the metal enrichment of the interstellar medium at the lowest 
metallicities. Thus we can assume that Pop III stars die instantaneously, 
while less massive stars are assumed to evolve according to their appropriate 
stellar lifetime. In this work, the lifetimes of stars of different mass and 
metallicity are computed according to the parametric form proposed by Raiteri 
et al. (1996).
As we will show in section \ref{sec:pop3}, our results are not sensitive
to the adopted values for the mass of Pop III stars and of the critical 
metallicity.

Finally, the galactic outflows driven by SN explosions, $dM_ {\rm ej,SN}/dt$,
are assumed to be proportional to the SN explosion rate ($\rm R_{\rm SN}$) 
divided for the escape velocity ($v_e$) squared:
\begin{equation}
  \frac{dm_{\rm ej,SN}}{dt} = \frac{2\epsilon_{w}E_{\rm SN} R_{\rm SN}}{v_e^2}
\end{equation}
where $\epsilon_w$ is the SN-wind efficiency and $E_{\rm SN}$ is the average
SN explosion energy (see SFS08 for further details).

\subsection{New features}
The new features implemented for the purposes of the present work are:
{\it (i)} the growth of the BH (via both gas accretion and 
mergers with companion BHs), {\it(ii)} the effect of the AGN feedback on the 
host galaxy,  {\it(iii)} the merger driven starbursts, i.e. bursts of star 
formation episodes enhanced by halos binary merging, and  {\it(iv)} the dust 
formation and evolution into the ISM. We will describe them in turn.

\subsubsection{BH growth}\label{sec:BHgrowth}

It is usually assumed that SMBHs are assembled by mergers with other BHs 
and/or by accretion of gas from the surrounding medium. 
Thus, the growth of a BH is determined by the seed BHs properties
(e.g. mass, redshift distribution and abundance), the efficiency of 
the accretion rate, the number of BH mergers and the dynamics of the coalesced BHs.
Theoretical studies suggest that a seed BH with mass in the range $[10^2 - 10^6] M_\odot$ 
can form either by the rapid collapse of Pop III stars (Heger \& Woosley 2002) 
or by the direct collapse of massive hot ($T>10^4 \rm{K}$) and dense 
gas clouds induced by gravitational instabilities (e.g. Bromm \& Loeb 2003; 
Begelman, Volonteri \& Rees 2006; Volonteri \& Natarajan 2009).

\begin{figure}
  \centering
  \includegraphics [width=7.0cm]{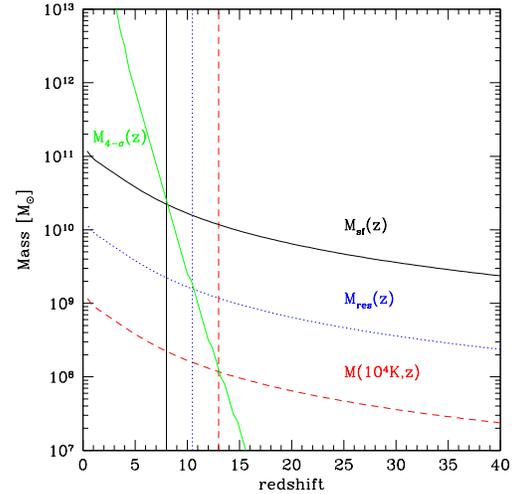}
  \caption{\small The threshold masses adopted in the code as a function of 
    redshift. $M(10^4 \rm K, z)$ is the mass of a halo with a virial 
    temperature $T_{\rm vir}=10^4$ K, $M_{\rm res}(z)=10 M(T_{\rm vir},z)$ is the 
    minimum halo mass in the merger tree simulations and 
    $M_{\rm sf}(z)$ is the minimum halo mass to form stars, 
    assumed to be $= 100 M(10^4 \rm K, z)$. 
    These masses are compared with the $4-\sigma$ mass density fluctuation 
    M$_{4-\sigma}$(z), green solid line. 
    When newly virialized halos, in the merger tree, are less massive than 
    M$_{4-\sigma}$(z) they do not host a seed black hole. The redshift at which
    halos are no more populated with seed BHs depends on the assumed threshold 
    as indicated by the vertical lines.
    For the adopted minimum halo mass to form stars, $M_{\rm sf}(z)$, we populate 
    halos down to redshift of about 8.} \label{fig:res}
\end{figure}

In this work we assign a BH seed, $M_{\rm seed} = 10^4 h^{-1} \rm{M_\odot}$ to all
the progenitor halos corresponding to density fluctuations higher than 
$4-\sigma$ ($M>M_{\rm 4-\sigma}$) that exceed the threshold mass required to form 
stars ($M>M_{\rm sf}$) for the first time in the merger tree. 
The evolution of these critical masses as a function of redshift is shown in 
Fig \ref{fig:res}. From this plot we can infer that the probability, $P$, 
that a halo with $M>M_{\rm sf}$ will host a seed BH depends on the redshift, 
other than on the specific formation history, and that $P \rightarrow 0$ for $z<8$.
Note that even assuming $P=1$ for all the star forming progenitors, the mass 
density parameter in seed BHs at redshift $z=6.4$ would be 
$\Omega_{\bullet} \sim 10^{-11}$, several orders of magnitude less than the SMBH 
density parameter estimated from nuclear BHs of nearby galaxies, 
$\Omega_{\rm SMBH} (z=0) = 2.9 \times 10^{-6}$ (Merloni \& Heinz 2008). 
Indeed, seed BHs grow via binary mergers and gas accretion.
The choice of the BH seed will be discussed in section \ref{sec:BHevo}.
Following Tanaka \& Haiman (2009), we assume that if the progenitor halos 
mass ratio, 
$\mu = M_{\rm s}/M_{\rm p}$ (where $M_{\rm s}$ is the mass of the secondary 
smaller halo and $M_{\rm p}$ is the mass of the primary larger halo), 
is $\mu > 1/20$ the BHs coalesce efficiently 
over the same timescale of their host halos, adding their masses linearly\footnote{This condition
is derived requiring that the halo merger timescale is shorter than the Hubble time (see Tanaka \& Haiman 2009 for more details).}. 
Conversely, if the mass ratio of the two progenitors is smaller than 1:20,
only the largest of two original BHs survives as a nuclear BH, the less 
massive one is assumed to end up as a satellite and its evolution along 
the merger tree is no longer followed.

It is known that the BH produced by a coalescence receive a gravitational 
recoil (kick velocities $\ge 100$~km/s) due to the net linear momentum 
accumulated by the asymmetric gravitational wave emission 
(Campanelli et al. 2006; Baker et al. 2006). As a consequence of this effect, 
the recoiling BH could be kicked out of its host halo or displaced in 
under-dense regions, thus reducing the gas accretion rate.
Tanaka \& Haiman (2009) have added in their Monte Carlo code an explicit 
calculation of the orbits of kicked BHs, and self-consistently include 
their corresponding time dependent accretion rate. 
We do not attempt to do a similar analysis here and we neglect this effect; 
note, however, that in our model we expect the number of BHs ejected out of 
their host halo to be modest since the escape velocities of halos with masses 
$\ge M_{\rm sf}(z)$ are $\ge 110$~km/s.  

In \textsc{GAMETE/QSOdust}, BHs are allowed to grow in mass by accreting gas 
from the surrounding medium. This process is triggered by gas-rich halo 
mergers and it is regulated both by star formation, which consumes gas by 
converting it into stars, and by AGN feedback, which reduces the amount 
of material available for accretion.
In our formulation, the central BH accretes gas at a rate given by 
$\dot{M}_{\rm accr}=min(\dot{M}_{\rm Edd}, \dot{M}_{\rm BHL})$, 
where $\dot{M}_{\rm Edd}$ is the Eddington rate,
\begin{equation}
\dot{M}_{\rm Edd} = \frac{4\pi G M_{\rm BH}m_{\rm p}}{\epsilon_{\rm r} \sigma_{\rm T} c} 
\label{eq:edd}
\end{equation}
\noindent
and $\dot{M}_{\rm BHL}$ is the Bondi-Hoyle-Lyttleton (BHL) accretion rate,
\begin{equation}
\dot{M}_{\rm BHL}=\frac{4 \pi \alpha G^2 M_{\rm BH}^2 \rho_{\rm gas}(r_{\rm A})}{c_{\rm s}^2}. 
\label{eq:bhl}
\end{equation} 
\noindent
In the above equations, $G$ is the gravitational constant, $m_{\rm p}$ is the 
proton mass $\sigma_{\rm T}$ is the Thomson cross section, $c$ is the light 
speed, $c_{\rm s}$ is the sound speed, $\rho_{\rm gas} (r_{\rm A})$ is the gas 
density evaluated at the radius of gravitational influence of the BH (or Bondi 
accretion radius), $r_{\rm A}=2G M_{\rm BH}/c_{\rm s}^2$, and $\alpha$ is a 
dimensionless parameter (see below); in eq.~\ref{eq:edd}, the parameter
$\epsilon_{\rm r}$ is the radiative efficiency, which determines the conversion 
efficiency of mass accretion into energy released as radiated luminosity, and 
it is fixed to 0.1 (Shakura \& Sunyaev, 1973). 

We adopt a gas density distribution described by a 
singular isothermal sphere (SIS) profile with a flat core: 
\begin{equation}
 \rho(R) = \frac{\rho_{\rm norm}}{1+(R/R_{\rm core})^2},
 \label{eq:gasprofile}
\end{equation}
\noindent
where $R_{\rm core} = 0.012 R_{\rm vir}$ is the size of the flat core, 
$R_{\rm vir}$ is the virial radius and $\rho_{\rm norm}$ is the normalization constant; 
the profile is normalized such that, at each time step, the total gas mass, $M_{\rm gas}$, 
is enclosed within the virial radius. 
The factor $\alpha$ which appears in eq.~\ref{eq:bhl} does not appear in the 
original analysis of Bondi \& Hoyle (1944) and Hoyle \& Lyttleton (1939), but has been 
introduced by Springel, Di Matteo \& Hernquist (2005) as a numerical correction factor, 
to compensate for the limitations of the numerical simulations. In fact, 
present simulations lack both the resolution and the physics to model the 
multiphase interstellar medium, and tend to strongly underestimate the 
Bondi-Hoyle accretion rate; values of $100 \le \alpha \le 300$ are commonly
adopted in the literature (Springel et al. 2005; Di Matteo et al. 2005, 2008;
Sijacki et al. 2007; see however Booth \& Schaye 2009 who parametrize $\alpha$ 
as a function of density). As it will be discussed in the following sections,
we also assume a constant $\alpha$ and fix its value in order to reproduce the
observed black hole mass of J1148.

\subsubsection{AGN feedback}

It is natural to expect that a quasar shining at, or close to, its limiting 
Eddington luminosity can generate a powerful galactic wind and eventually 
terminate the accretion process that feeds it. If the energy in the
outflow liberates as much energy as the binding energy of the gas in a dynamical
time, the feedback from the accreting BH can be self-regulated 
(Silk \& Rees 1998; Wyithe \& Loeb 2003).
Hydrodynamical simulations suggest that BH accreting at high rates release 
enough energy to drive galactic scale outflows of gas (Springel et al. 2005; 
Di Matteo et al. 2005; Ciotti, Ostriker \& Proga 2009, 2010). 
The mechanism by which the energy released by the accreting BH is coupled to 
the surrounding medium is currently under debate (see Ciotti et al. 2010 and 
references therein). The energy transfer rate is commonly parametrized as 
(e.g. Springel et al. 2005),
\begin{equation}
\dot{E}_{\rm fbk}=\epsilon_{\rm w,\textsc{agn}} \epsilon_{\rm r} \dot{M}_{\rm accr}c^2, 
\label{eq:power}
\end{equation}
\noindent
where the parameter $\epsilon_{\rm w,\textsc{agn}}$ is the coupling efficiency and it
is usually assumed to be a free parameter of the simulation, independent of the 
environment and gas properties. 

Note that recently, Sijacki et al. (2007) have differentiated between a 
'quasar mode' and a 'radio mode' feedback, depending on the accretion rate. 
In particular, BH-powered winds are related to the luminous quasar activity 
and are efficient for high accretion rates. 
They are expected to be more powerful than starburst-driven winds and should 
be able to sweep away a significant amount of gas from the deep potential well 
of their massive host galaxies.
Once the obscuring material is swept away from the center of the system, the
quasar enters its visible phase in which it outshines the entire galaxy (quasar
 mode).
In the low-accretion rate regime, which is subdominant in terms of black hole
mass growth, the feedback is characterized by AGN-driven bubbles; this 
regime can be identiﬁed with radio galaxies in clusters (radio mode).
The BH accretion rate which signs the transition between these two modes
is usually assumed to be 
$\dot{m}_{\rm crit}=(\dot{M}_{\rm accr}/\dot{M}_{\rm Edd})=10^{-2}$
(Sijacki et al. 2007); thus, since we are not interested here in the 
radio-mode regime, we only activate the BH feedback when the accretion rate is 
super-critical. 

When the above condition is verified, we estimate the amount of gas mass 
ejected out of the host halo per unit time as,
\begin{equation}
\frac{dM_{\rm ej,AGN}}{dt} = 2 \epsilon_{\rm w,\textsc{agn}} \epsilon_{\rm r} \big(\frac{c}{v_{\rm e}} \big)^{2} \dot{M}_{\rm accr}  
\label{eq:agnfdb}
\end{equation} 
\noindent
where $M_{\rm ej,AGN}$ is the mass of ejected gas and 
$v_{\rm e}=\big(\frac{GM}{R_{\rm vir}} \big)^{1/2}$ is the escape velocity of the 
galaxy. 

\subsubsection{Bursting star formation mode}

The star formation history (SFH) of the QSO host at redshift $z=6.4$ is the 
result of the SFHs of all its progenitor halos. 
There are observational and theoretical indications that support the idea that 
galaxy mergers have a dramatic effect on their morphological evolution and 
star formation history (see e.g. Woods \& Geller 2007; Ellison et al. 2010;
Mihos \& Hernquist 1996; Cox et al. 2008 and references therein). 
In particular, enhanced 
central star formation in interacting galaxies result from 
gaseous inflows that occur when the gas loses angular momentum through 
gravitational tidal torques produced primarily by the non-axisymmetric 
structure induced by the companion galaxies. 
Indeed, there is mounting evidence that near equal
mass mergers (with mass ratios greater that 1:3) triggers the most vigorous 
star-forming galaxies in the local universe, the ULIRGs (see e.g. Genzel et al.
 2001; Tacconi et al. 2002; Dasyra et al. 2006). 
Similarly, there is growing observational evidence for interaction-driven 
nuclear activity (see the recent review by Shankar 2009).

Based on a set of numerical simulations, Cox et al. (2008) have studied the 
effects of the galaxy mass ratio onto merger-driven starbursts. 
They find that while mergers between nearly equal mass galaxies produce the 
most intense bursts of star formation (major mergers), minor mergers produce 
relatively little enhancement with respect to the quiescent mode of star 
formation. These simulations have been calibrated to reproduce the observed 
properties of local disk galaxies and may not be immediately extended to 
higher redshifts. The resulting fitting formula have been implemented in 
semi-analytic models of galaxy evolution which successfully reproduce a number 
of observations at $z \sim 0$ (Somerville et al. 2008).

Here we are interested in exploring the effects of different SFHs on the 
chemical properties of QSOs host galaxies. 
Therefore we investigate both quiescent and bursting models: in the first 
class of models, star formation occurs only in a quiescent mode which is 
independent on the galaxy mergers mass ratio; conversely, in the second class 
of models an additional contribution to the stellar mass comes from galaxy 
major mergers. In order to explore a wide parameter space, we 
parametrize the global redshift dependent efficiency of star formation per unit
time that appears in eq.~\ref{eq:sfr} as,
\begin{equation}
f_{\ast}(z)=(\epsilon_{\rm quies}+\epsilon_{\rm burst})/t_{\rm dyn}(z) 
\label{eq:sfeff}
\end{equation}
\noindent
where $t_{\rm dyn}(z) = R_{\rm vir}/v_{\rm e}$ is the dynamical time, 
$\epsilon_{\rm quies}$ is the constant star formation efficiency during the 
quiescent phase of the evolution and $\epsilon_{\rm burst}$ is the starburst 
efficiency, expressed as a normalized Gaussian distribution of the mass ratio,
\begin{equation}
\epsilon_{\rm burst}=\frac{1}{\sqrt{2\pi}\sigma_{\rm burst}}e^{-(\mu-\mu_{\rm crit})^2/2\sigma_{\rm burst}^2}. 
\label{eq:bursteff}
\end{equation} 
\noindent
The starburst efficiency peaks at the critical mass ratio $\mu_{\rm crit}=1$ and 
the dispersion parameter controls the intensity of the burst for a given mass 
ratio. It is important to stress that since the merger rate and the mass ratio 
of merging halos depend on the hierarchical merger history, variations of the 
parameter $\sigma_{\rm burst}$ allow to explore a wide range of SFHs, with 
lower values of $\sigma_{\rm burst}$ corresponding to fewer intense bursts over 
the history of the galaxy; as $\sigma_{\rm burst}$ decreases, the SFHs will be 
characterized by more frequent and less intense bursts, progressively 
approaching the quiescent SF mode. 
For our test cases, $\sigma_{\rm burst}= 0.05, 0.1, 0.25$ corresponding to
maximum mergers efficiencies (if $\mu=\mu_{\rm crit}$) of 
$\epsilon_{\rm burst}\simeq 8, 4, 1.6$, respectively.

\subsubsection{The evolution of dust}
\label{sec:dustevo}
We compute the evolution of dust under the following assumptions:
{\it (i)} both SNe and AGB stars contribute to dust production on 
different timescales according to their stellar progenitors lifetimes;
{\it (ii)} there is no dust infall or outflow to/from the ISM, i.e. the 
infalling gas is dust free: only dust grain metals which are returned to the 
gas-phase ISM through destruction by interstellar shocks (i.e. dust which has 
been shocked by SNe) can be ejected out of the galaxy;
{\it (iii)} a fraction $f_{mc}$ of the dust mass is enclosed in dense MCs, 
$M_d^{mc}(t) = f_{mc} M_d(t)$; this component can grow through mass accretion of 
metals onto pre-existing grains in a time scale $\tau_{acc}$; 
{\it (iv)} the remaining dust component, $M_d^{diff}(t)=(1-f_{mc})M_d(t)$, is diffused 
in the ISM and can be destroyed by interstellar shocks in a time scale $\tau_{d}$.  

The destruction and accretion time scales depend on the properties of the ISM 
and on the cycling times between
the cloud (cold and dense) and intercloud (warm/hot and diffuse) phases (e.g. 
Tielens et al. 1998, 2005). Thus, a multiphase description of the ISM
would be required to properly follow the evolution of the two dust components, 
$M_d^{mc}$ and $M_d^{diff}$. 
Here, we make the simplifing assumption that a fixed fraction 
$f_{mc} \sim 1/2$ (see e.g. Dwek 1998) of the total dust particles 
are located into MCs and thus can serve as growth centers for the 
accretion of icy mantels (e.g. Draine 1990).

Following Hirashita (1999), we model the dust accretion timescale $\tau_{acc}$
as: 
\begin{equation}
  \tau_{acc}=\frac{\tau_{mc}}{X_{cold} [1-f_{dep}(t)]}, \label{eq:acctime}
\label{eq:tauacc}
\end{equation}
where, $\tau_{mc} = 3 \times 10^7$~yr is the typical lifetime of a MC 
(e.g. Tielens 1998), $X_{cold}$ is the fraction of cold gas into the ISM, which
is set to unity in our one-phase scenario. Finally, $f_{dep}(t)$ is the 
depletion factor, i.e. the ratio between the mass of dust and the total mass
in metals $M_d(t)/M_Z(t)$ which is computed at each time in each halo of 
the simulation. 
Only metals which are not already condensed into dust grains
(i.e. gas-phase metals) are available for accretion.
Typical values of this timescale are in the range $(10^7-10^8)$ yr (see Fig.~\ref{fig:timescales}).

The destruction timescale, $\tau_{d}$, i.e. the lifetime of  
dust grains against destruction by thermal sputtering in high-velocity 
($v>150$ km s$^{-1}$) SN shocks is given by (e.g. Dwek \& Scalo 1980; 
McKee 1989) :
\begin{equation}
\tau_{d}=\frac{M_{\rm ISM}(t)}{\epsilon_d M_{swept}R'_{\rm SN}},\label{eq:destime}
\end{equation}
\noindent
where $\epsilon_d$ is the dust destruction efficiency, 
$M_{\rm swept}$ is the effective ISM mass that is completely cleared of dust by 
a single SN remnant and  $R'_{\rm SN}=f_{\rm SN} R_{\rm SN}$ is the 
effective SN rate for grains destruction, which takes into account that not 
all the SNe which interact with the ISM are efficient in destroying 
dust\footnote{The most massive stars 
exploding in an association will be very effective in destroying dust while 
sweeping up matter into a shell. Subsequent supernovae will interact 
predominantly with the gas inside the shell and the shell itself and, 
therefore, do not contribute much to dust destruction. 
The expansion of the supershell is too slow to destroy much dust
(e.g. Tielens 2005).} (see e.g. McKee 1989; Tielens 2005); 
for PISNe $f_{SN}=1$ and for core collapse SNe $f_{SN}\sim 0.15$ 
(i.e. 15\% of all SNe in a stellar generation are efficient). 
Following McKee (1989), we compute the mass of the interstellar matter shocked 
to a velocity $v_{sh}$ in the Sedov-Taylor stage, in a homogeneous ISM, as:
\begin{equation}
  M_{swept}=6800 M_\odot \frac{(E_{\rm SN}/10^{51} \rm erg)}{(v_{sh}/100 \rm km s^{-1})^2},
\end{equation}
where $v_{sh}\sim 200$ km s$^{-1}$ is the (minimum) non-radiative shock velocity
and $E_{SN}$ is the average supernova explosion energy, assumed to be 
$2.7\times 10^{52}$ erg for PISNe and $1.2\times 10^{51}$ erg for core collapse
SNe.
We adopt the dust destruction efficiencies from Nozawa et
al. (2006) for PISNe and from Jones et al. (1996) for SNe. Assuming a reference
value for the ISM density of $n_{\rm ISM}=1$cm$^{-3}$ we get $\epsilon_d \sim 0.6 (0.48)$ for PISNe 
(SNe). 
With these parameters, we find that for conditions which apply to the Milky 
Way Galaxy ($M_{ISM}\sim 5\times 10^9$ M$_\odot$, $R'_{\rm SN}\sim 7.5\times 
10^{-3}$ yr$^{-1}$, McKee 1989), typical grain lifetimes are $\sim 0.6$ Gyr, 
in agreement with Jones et al. (1996).

\section{The chemical evolution model with dust}\label{ref:chemevo}
In this section we  describe the chemical evolution model implemented in the 
code \textsc{GAMETE/QSOdust}. 

We will indicate with $M_{\rm ISM}$ the total mass in the interstellar
medium ($M_{\rm gas}+M_{d}$) and with $M_Z$ the total mass of metals (diffused in 
the gas phase and condensed into dust grains). 
At any given time,
the mass of {\it gas phase} metals is calculated as the difference between $M_Z$ and $M_d$ 
(with proper stoichiometric coefficients when we are interested in a 
particular gas phase element).
The variations of the total mass of gas, stars, metals and dust is followed 
in each progenitor halo of the hierarchical tree solving the following set of
differential equations:
\begin{equation}
  \frac{dM_{\ast}(t)}{dt}  =  \rmn{SFR}(t) - \frac{dR(t)}{dt},
\end{equation}
\begin{eqnarray}
  \frac{dM_{\rm ISM}(t)}{dt} & = & - \rmn{SFR}(t) + \frac{dR(t)}{dt}+ 
  \frac{dM_{inf}(t)}{dt} - \frac{dM_{ej}(t)}{dt}  \nonumber \\
  & & - (1-\epsilon_r)\frac{dM_{accr}(t)}{dt}
\end{eqnarray}
\begin{eqnarray}
  \frac{dM_{Z}(t)}{dt}  =  - Z_{\rm ISM}(t) \rmn{SFR}(t) + \frac{dY_{Z}(t)}{dt}+
   Z_{\rm vir}(t)\frac{dM_{inf}(t)}{dt} \nonumber \\
   - Z_{ISM}(t)\frac{dM_{ej}(t)}{dt}   
   - Z_{\rm ISM}(t)(1-\epsilon_r)\frac{dM_{accr}(t)}{dt}\label{eq:met}
\end{eqnarray}
\begin{eqnarray}
  \frac{dM_{d}(t)}{dt} & = & - Z_d(t) \rmn{SFR}(t) + \frac{dY_{d}(t)}{dt} 
   - \frac{M_{d}^{diff}(t)}{\tau_{d}}+\frac{M_{d}^{mc}(t)}{\tau_{acc}}\nonumber\\
   & & - Z_d(t)(1-\epsilon_r)\frac{dM_{accr}}{dt} \label{eq:dust}
\end{eqnarray}
where $M_{\ast}$ and $M_d$ are the total mass of stars and dust,
$Z_{\rm ISM}(t)=M_{Z}(t)/M_{\rm ISM}(t)$ is the ISM metallicity, 
$Z_{d}(t)=M_{d}(t)/M_{\rm ISM}(t)$ is the total dust abundance in the
ISM and $Z_{\rm vir}$ is the metallicity of the infalling gas, namely of the hot
gas at the virialization epoch. Its value corresponds to the metallicity of 
the external medium, that is the ratio between the total mass of metals 
ejected by star forming haloes and the mass of the diffused gas, not 
enclosed in collapsed objects (see SFS08 for details). 

The new features of the code are summarized by: 
{\it (i)} the evolution of dust described by eq. \ref{eq:dust};
{\it (ii)} the term $(1-\epsilon_r)dM_{accr}/dt$ which accounts for the fraction,
($1-\epsilon_r$), of the gas (metals and dust), which fuels BH growth via 
accretion;
{\it (iii)} the gas ejection rate $dM_{ej}(t)/dt = dM_{ej,SN}(t)/dt + 
dM_{ej,AGN}(t)/dt$ which includes both SNe and AGN feedback. 

Finally, the terms $dR/dt$ and $dY_{Z}/dt$ and $dY_d/dt$ are the rates at which
gas, heavy elements and dust are returned to the ISM. 
The equations and grids of yields used to compute the gas and metal return 
fraction are the same as in SFS08 and V09: 
van den Hoeck \& Groenewegen (1997) for AGB stars with initial metallicities 
$Z=(5 \times 10^{-2}, 0.2, 1) Z_{\odot}$ and masses $(1-8) \rm M_{\odot}$; Woosley 
\& Weaver (1995) for SNe with initial masses $(12 - 40) \rm M_{\odot}$ at
 metallicities $Z=(0, 10^{-4}, 10^{-2}, 10^{-1}, 1)$ Z$_{\odot}$; Heger \& 
Woosley (2002) for the PISNe assuming a reference progenitor mass of 
200 M$_{\odot}$.
The dust injection rate $dY_d(t)/dt$ is computed as in eq.~(12) of V09, 
including contributions from AGB stars, SNe and PISNe. We use the grids of
Zhukovska, Gail \& Trieloff (2008) for AGB stars with $0.8 M_{\odot} \lesssim m 
\lesssim 8 M_{\odot}$ at different metallicities; for SNe with progenitor mass 
of $(12 - 40) \rm M_{\odot}$\footnote{Dust and metal yields 
from Pop II/I stars with masses in the range 
$41-100 \rm M_\odot$ are assumed to be zero. 
Indeed, it has been shown (e.g. Dwek et al. 1998; 
Zhukovska et al. 2008 and reference therein) that such massive stars 
provide only a minor contribution to the total dust in the ISM with respect to 
that of AGB stars and core-collapse SNe.}
 and metallicities $Z=(0, 10^{-4}, 10^{-2}, 10^{-1}, 1)$Z$_{\odot}$,
we take the model by Bianchi \& Schneider (2007) where grain condensations starts 
from seed clusters made of $\cal{N} \geq$ 2 monomers with a sticking coefficient of $\alpha=1$.
We also assume that SNe explode in an ISM with a density of $\rho_{\rm ISM}=10^{-24}$g cm$^{-3}$,
so that only 7\% of the initial dust formed in the ejecta survives the passage of the reverse shock.
The resulting dust masses ($10^{-3}-10^{-1}$ M$_\odot$ depending on the 
progenitor star initial mass and metallicity) are in 
agreement with the values inferred from observations of young SN remnants 
(e.g. Hines et al. 2004; Rho et al. 2008; Sandstrom et al. 2008; Rho et al. 
2009). Finally, for PISNe we adopt the dust yields computed by Schneider, Ferrara \& Salvaterra (2004) for PISNe. 
As in V09 we extrapolate from the dust mass produced by the largest 
available AGB model (7 $M_{\odot}$) and by the smallest SN progenitor 
(12 $M_{\odot}$) to account for metal and dust production in the mass range
$(8-11)$ M$_{\odot}$ for which the dust yields are not yet available in the 
literature. The grids of Zhukovska et al. (2008) are not computed at 
metallicity $Z <5 \times 10^{-2} Z_{\odot}$; thus we have assumed that the dust mass produced
by AGB at lower metallicities is the same as that at 
$Z =5 \times 10^{-2} Z_{\odot}$\footnote{Note that this approximation may hold 
for carbon dust but silicate dust production is likely to be significantly 
lower at low metallicity (see V09).}. 

\section{Model parameters}
Our main aim is to investigate the effect of different SFHs in modeling the chemical 
properties of the host galaxy of QSO J1148. The free parameters
which determine the shape of the SFH are:
\begin{itemize}
    \item the star formation efficiency, $f_\ast(z)$, which is 
      determined by the choice of $\epsilon_{\rm quies}$ and $\sigma_{\rm burst}$ 
      (see eq. \ref{eq:sfeff});
    \item the efficiency of the AGN-driven wind, $\epsilon_{w,AGN}$ (see eq. 
      \ref{eq:agnfdb});
    \item the efficiency of gas accretion onto the BHs, $\alpha$ (see eq.
      \ref{eq:bhl}).
\end{itemize} 
 
The parameters which define the models considered in this work are 
summarized in Table \ref{tab:models} where letters and numbers given in the 
first column identify each model. 

These have been selected in order to reproduce the BH and total gas mass
observed at $z=6.4$. Given the uncertainties in the derivation of the stellar mass discussed in 
section \ref{sec:hostprop}, we have not constrained evolutionary 
models in terms of the predicted final stellar mass. 
Indeed, by varying the global star formation efficiency, $f_\ast(z)$, we 
explore the dependence of the final mass of metals and dust on the stellar 
mass and SFH, considering models which predicts a stellar mass 
compatible with the value inferred by the observations ($3.9 \times 10^{10} M_{\odot}$,
 Walter et al. 2004) as well as models with higher stellar masses, closer to 
the values implied by the present-day $M_{\rm BH}$-$M_\ast$ relation 
($\sim 10^{12} M_{\odot}$, see Fig.~\ref{fig:bhrel}).
The rate at which gas is converted into stars in quiescent (Q)
models is regulated by the choice of the efficiency 
$\epsilon_{\rm quies}$ given in the second column of the Table
(for these models $\epsilon_{\rm burst}=0$ in eq. \ref{eq:sfeff}). 
In bursted (B) models the frequency and amplitude of starbursts are 
regulated by the parameter $\sigma_{\rm burst}$. The dependence on the choice of
this latter parameter is investigated in Appendix A.

It has been shown that assuming an AGN wind efficiency 
$\epsilon_{\rm w,\textsc{agn}}=5\%$ many observed properties of elliptical 
galaxies as well as the observed $M_{\rm BH}-\sigma$ correlation can be 
reproduced (see the discussion in Hopkins et al. 2006). 
Here, we follow a similar approach:
in all the models presented here, we choose the parameters $\alpha$ 
and $\epsilon_{w,AGN}$ in order to reproduce the observed BH mass and to 
mimic a powerful AGN-driven wind which is able to sweep away enough gas from 
the host galaxy to reproduce the observed gas content and make the quasar 
optically visible, in its active phase, at $z = 6.4$.
We set the parameter $\alpha$ in the range $180-350$, to allow enough gas 
accretion to fuel the observed SMBH and the wind efficiency 
$\epsilon_{w,\textsc{agn}}$, is chosen to regulate simultaneously the BH 
accretion, star formation and gas ejection (to produce a global gas mass in 
agreement with the observational lower limit of Walter et al. 2004, see 
section \ref{sec:hostprop}).

Strictly speaking, the parameters listed in Table 2 are not independent.
With higher star formation efficiencies, $f_\ast$, less gas will be available 
for BH growth, thus requiring a higher BH accretion efficiency, $\alpha$, in 
order to reproduce the observed BH mass.
On the other hand, the higher is BH accretion rate the stronger is the 
AGN wind (and the earlier it affects the global gas content), thus we need 
to set properly the coupling efficiency, $\epsilon_{w,\textsc{agn}}$, to drive a 
gas outflow which expels the required amount of gas, but avoids a too early 
decline of the SFR and BH accretion. 
Therefore, as it has emerged from the extensive 
exploration of the parameter space out of which the present models have been
selected, the set of parameters which characterize each model is not unique. 
Given this degeneracy, we keep the value 
$\epsilon_{w,\textsc{agn}}=5\times 10^{-3}$ fixed as we vary 
$\alpha$, in order to easily compare the AGN feedback effects in different 
models and control the combined effect of BH accretion and 
AGN-driven gas outflow, in shaping the SFH. 
Unfortunately, the AGN-driven outflows are very 
difficult to observe if the quasar is in its active phase. As in the case of
J1148, optically bright quasars outshine their host galaxy and ionize the 
cold gas which is usually a tracer of the outflows. Thus, there are no 
observational constraints on the feedback process in such peculiar high 
redshift objects.

\begin{table}
  \begin{center}
    \begin{tabular}{|c|c|c|c|c|c|}
      \hline
      {\bf model} & {\bf $\epsilon_{\rm quies}$} & {\bf $\sigma_{\rm burst}$} & {\bf $\epsilon_{\rm burst,max}$} & {\bf $\alpha$} & {\bf $\epsilon_{w,\textsc{agn}}$}\\
      \hline   \hline
      Q1    & 0.045 &  -  & - &   190   &  $5\times10^{-3}$ \\
      Q2    & 0.1   &  -  & - &   180   &  $5\times10^{-3}$ \\
      Q3    & 1.0   &  -  & - &   350   &  $5\times10^{-3}$ \\
            &       &     &   &         &                  \\
      B1    & $6.7\times 10^{-3}$ &  0.05  & 0.53 &  180 &  $5\times10^{-3}$ \\
      B2    & 0.02  &  0.05   & 1.6  &   200   &  $5\times10^{-3}$ \\
      B3    & 0.1   &  0.05   & 8.0  &   200   &  $5\times10^{-3}$ \\
      \hline
    \end{tabular}
  \end{center}
  \caption{\small 
    Model parameters. The models presented 
    in this work are divided in two classes on the basis of the simulated SFH:
    quiescent only (Q) and bursting (B) star formation models. The numbers in 
    the first column identify the efficiencies of star formation in quiescent 
    phases. 
    The second column provide the value of $\sigma_{\rm burst}$ which determines 
    the number, redshift distribution and intensity of the starbursts 
    (B-models only). 
    The maximum starburst efficiency, $\epsilon_{\rm burst,max}$ is given in the 
    third column (B-models only). 
    Finally, the efficiency of gas accretion into the central BH, $\alpha$, 
    and the efficiency of AGN-driven wind, $\epsilon_{w,AGN}$, are shown in the 
    last two columns.}
  \label{tab:models}
\end{table}

\section{Results}
In this section we present the results of the semi-analytical hierarchical model
\textsc{GAMETE/QSOdust}. We have performed 50 random 
realizations of the quasar hierarchical assembly (as explained in section~3.1).
Each of these merger histories allows us to predict a specific 
evolutionary path for the host galaxy chemical properties. 
However, to make the results and conclusions of this work independent of the 
hierarchical formation history, we will present the 
predicted evolution of the BH and galaxy properties averaging over all the 50 
merger tree realizations. 

\subsection{BH evolution}\label{sec:BHevo}
\begin{figure}
  \centering 
  \includegraphics[width=8.6cm]{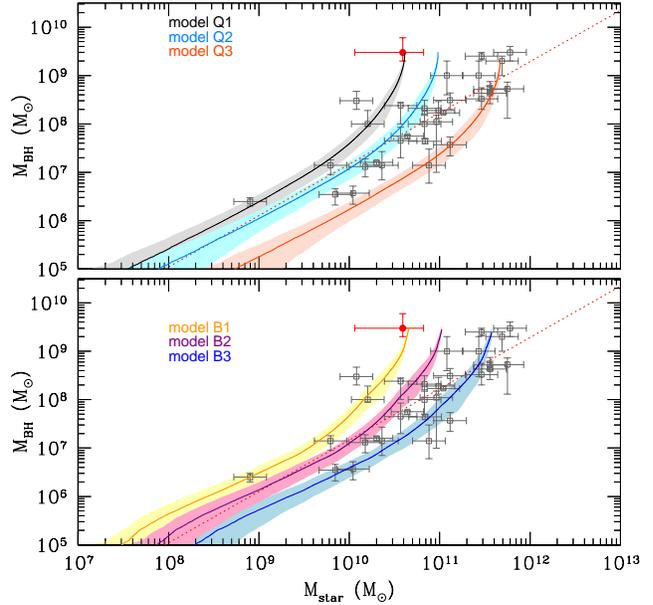}
  \caption{\small
    The evolution of the BH mass as a 
    function of the stellar mass predicted by quiescent models (upper 
    panel) and bursted models (lower panel). 
    Open squared data points with error bars are from Marconi \& Hunt (2003, 
    Table 1 in their work) and represent the $M_{\rm BH}-M_\ast$ relation 
    observed in the local Universe, with the empirical fit
    $M_{\rm BH}/M_{\ast} \sim 0.002$ (dotted line) provided by the same authors. 
    The filled circle shows the J1148 BH (Willot et al. 2003) and 
    stellar masses (Walter et al. 2004) inferred from observations.
    For all models, solid lines are averages over 50 merger tree realizations 
    and shaded areas represent $1\sigma$ dispersion.} 
  \label{fig:bhrel}
\end{figure}

As introduced in section \ref{sec:hostprop}, we investigate the effects of the 
assumed SFH considering either models which predict stellar masses in 
agreement with the J1148 position on the $M_{\rm BH}-M_\ast$ plane and with what 
would be implied if the system were to follow the observed local relation, 
taking into account the observational dispersion.
Such models are shown in Fig.~\ref{fig:bhrel} which depicts the evolution onto the $M_{\rm BH}-M_\ast$
plane predicted by quiescent models Q1, Q2, Q3 (upper panel) and bursted
models B1, B2, B3 (lower panel). 
By construction, all models reproduce the observed J1148 SMBH mass but predict 
stellar masses increasing with $f_\ast$: low-$f_\ast$ models (Q1
and B1) reproduce the $M_\ast$ inferred from 
observations; high-$f_\ast$ models (Q3 and B3) 
predict a final $M_{\rm BH}-M_\ast$ correlation which is in 
agreement with the local observations (squared data points). 
We note that, despite the different final stellar masses, all the models
tend to approach the final value from the bottom.
In other words we find that in most of the models the stellar bulge 
appears to evolve faster than the central BH  ($M_\ast > 10^{10}$ M$_\odot$
always for $M_{BH}\geq 10^{8}$ M$_\odot$). 
This differs from the recent study by Lamastra et al. (2010) who predict an {\it evolution} 
with redshift of the $M_{\rm BH}-M_\ast$, with higher-$z$ galaxies forming their 
nuclear BH faster than their stellar mass. This may be due to the different physical 
description of BH accretion and feedback which, in the Lamastra et al. (2010)
study, tend to enhance the accretion efficiency at high redshift.

\begin{figure}
  \centering
  \includegraphics[width=8.8cm,height=7.0cm]{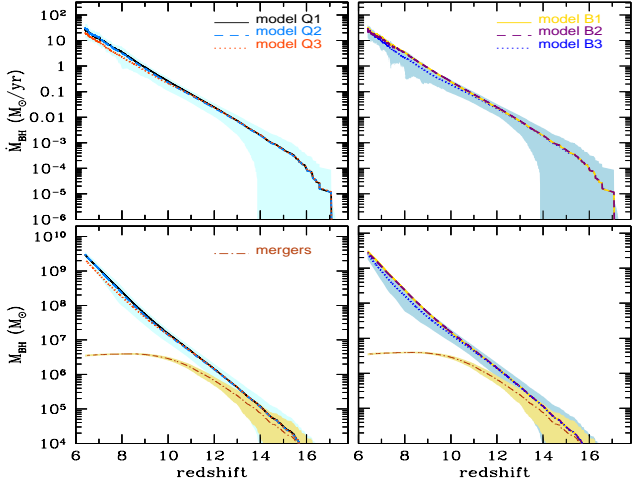}
  \caption{\small
    The black hole accretion rate, $\dot{M}_{\rm BH}$, and black hole mass 
    $M_{\rm BH}$ 
    evolution as a function of redshift (upper and lower panels, respectively), 
    in models Q (left panels) and B (right panels). 
    The dot-dashed curves in the bottom panels represent the contribution to 
    the BH mass by BH mergers.}
    \label{fig:bhevo}
\end{figure}

For each model and at each redshift, we 
compute the \emph{total} BH mass and accreted gas mass as the sum of all 
nuclear BHs and accretion rates hosted in each 
progenitor halo\footnote{Note that
at each redshift/timestep of the simulation, each BH and its host halo are the
progenitors of the final SMBH and its host galaxy which, by construction, will 
form at redshift $z=6.4$. 
Therefore, at each redshift the sum of all single-halo BH
 masses, as well as of the other quantities, such as the DM halo mass, the gas,
 stars, metals and dust masses, are representative of the properties of the 
final halo at early times ($z>6.4$).}.  
These are shown in Fig.~\ref{fig:bhevo}.
Due to the strong correlation among the parameters discussed 
above, the BH accretion rate (upper panels) and BH mass (lower panels) 
evolution have a negligible dependence on the particular model.
At redshift $z=6.4$, we predict that J1148 has a BH mass of $\sim 3\times 
10^9$ M$_\odot$ and accretes up to $\sim$ (20-30) $\rm M_\odot$/yr of gas; 
in units of the Eddington rate, the predicted accretion rates are 
$\dot{M}_{\rm BH}/\dot{M}_{\rm Edd} \sim (0.3-0.45)$ for all the models 
investigated.
The thin line in the BH mass evolution (bottom panels) represents
the final BH mass that would be produced only through BH mergers. 
It is clear from the figures that even if BH mergers are the main 
drivers for the early growth of the BH mass at $z>11$, their contribution 
to the final BH mass at $z=6.4$ is less than 1\%.

At redshift $z>11$ the dependence on the particular hierarchical formation 
history is stronger than at lower redshifts, as indicated by the larger shaded 
regions. Thus, differences among different realizations can influence 
only the early evolution of nuclear BHs. At $z\lesssim 11$ the main BH growth
mechanism is gas accretion.

\begin{figure}
  \centering
  \includegraphics[width=8.8cm, height=7cm]{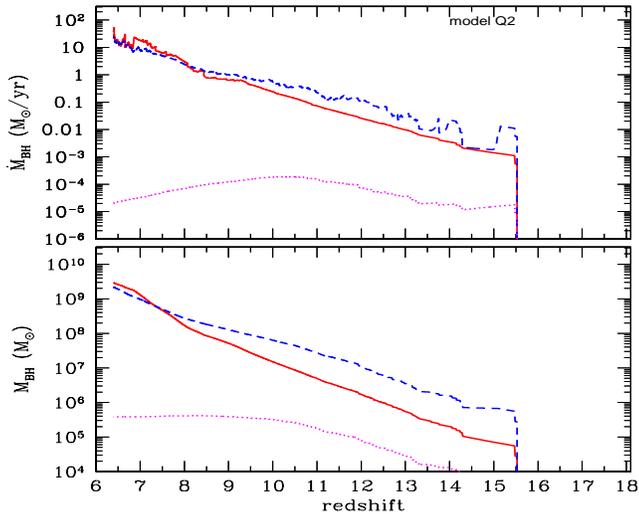}
  \caption{\small
    The black hole accretion rate, $\dot{M}_{\rm BH}$, and black hole mass, $M_{\rm BH}$, evolution as a
    function of redshift in model Q2 for different seed BH 
    masses. The magenta (dotted), red (solid) and blue (dashed) 
    lines show the results for a single merger tree realization assuming 
    seed BH masses of $10^3 h^{-1} M_{\odot}$, $10^4 h^{-1} M_\odot$, and 
    $10^5 h^{-1} \rm M_\odot$, respectively.}
    \label{fig:bhevoseed}
\end{figure}

In Fig. \ref{fig:bhevoseed} we explore, for a single merger history of model Q2,
the dependence of the evolution of the BH mass on the assumed seed BH. 
We find that models with $M_{\rm seed}\leq 10^3 h^{-1}$ cannot reproduce the 
observed BH mass without increasing the accretion efficiency ($\alpha$).
On the other hand, for seed BHs with mass larger than $\sim 10^4 \rm M_\odot$, 
BH accretion rate and BH mass at the final redshift $z=6.4$ do not depend much 
on the adopted seed. 
This is due to the interplay between Eddington limited BH accretion and AGN 
feedback processes which efficiently regulate BH growth. 
These results justify the seed BH mass adopted in the 
models (see section \ref{sec:BHgrowth}. 

\subsection{Star formation histories}\label{sec:sfr}
\begin{figure}
  \centering
  \includegraphics[width=8.8cm, height=7cm]{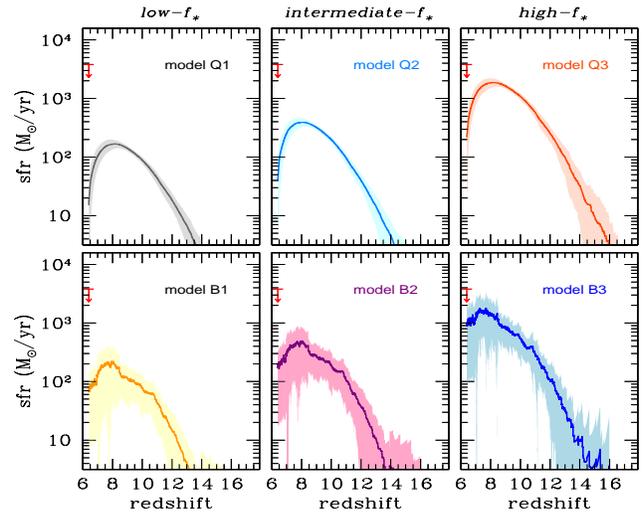}
  \caption{\small
    The star formation histories of the host galaxy of the quasar J1148 in 
    quiescent models, Q1, Q2, Q3 (upper panels, from left to right) and 
    in bursted models, B1, B2, B3 (lower panels, 
    from left to right). The arrow indicates the upper limit SFR inferred
    from the observation of J1148.
    The SFHs shown in this figure are obtained averaging 
    over 50 random halo merger tree realizations and are shown 
    together with the $1\sigma$ error represented by the shaded areas.}
  \label{fig:sfr}
\end{figure}

The different SFHs produced by quiescent (Q1, Q2 and Q3) and bursted (B1, B2 
and B3) SFH models are shown in Fig. \ref{fig:sfr} (upper and lower panels, 
respectively), with increasing global star formation efficiencies, from
{\it low-} to {\it high-$f_\ast$}.

As it can be seen from the figure all SFHs peak at redshift $\sim 8$, with 
peak amplitudes increasing with $f_\ast$. 
The subsequent decrease is due to the effect of AGN feedback triggered by a 
large enough gas accretion rate onto the central BH.
The galactic scale wind expels gas from the halo (see Fig. \ref{fig:chemevo}), 
reducing the SF rate by a factor of $\sim 2-10$. The suppression of the SF rate
is stronger in Q-models where the rate of star formation only depends on the 
available mass of gas at each redshift and does not depend on the merging 
halos mass ratio and redshift distribution. 
Indeed, such models are only mildly affected by the variation among the 
different merger tree realizations, as also demonstrated by 
the narrow $1\sigma$ dispersion.

Conversely, in the B-models, the available gas mass at each redshift and thus
the shape of the SFH, is determined by the enhanced starburst efficiency which 
depends on the parameter $\sigma_{\rm burst}$. 
The number, intensity and redshift distribution of the 
starbursts strongly reflect the hierarchical merger history of the hosting DM
halo. 
In fact, for the same set of parameters, each model shows a wide dispersion
among the results obtained by individual merger histories, with their
specific redshift distribution of major mergers which trigger the starbursts.
The individual SFHs show very different features, such as the onset of 
starbursts 
or the redshift at which the last major merger occurs.
This is reflected in the large $1\sigma$ dispersions which characterize B-models.

We note that, only model B2 and high-$f_\ast$ models
predict average SFRs at redshift $z=6.4$ in the range $(200-900)$ M$_\odot/$yr, in
agreement with what inferred from observations of J1148, as discussed in section \ref{sec:obs}.

\subsection{Chemical evolution of the host galaxy}\label{sec:chemevo1}

\begin{figure*}
  \begin{center}
    \includegraphics{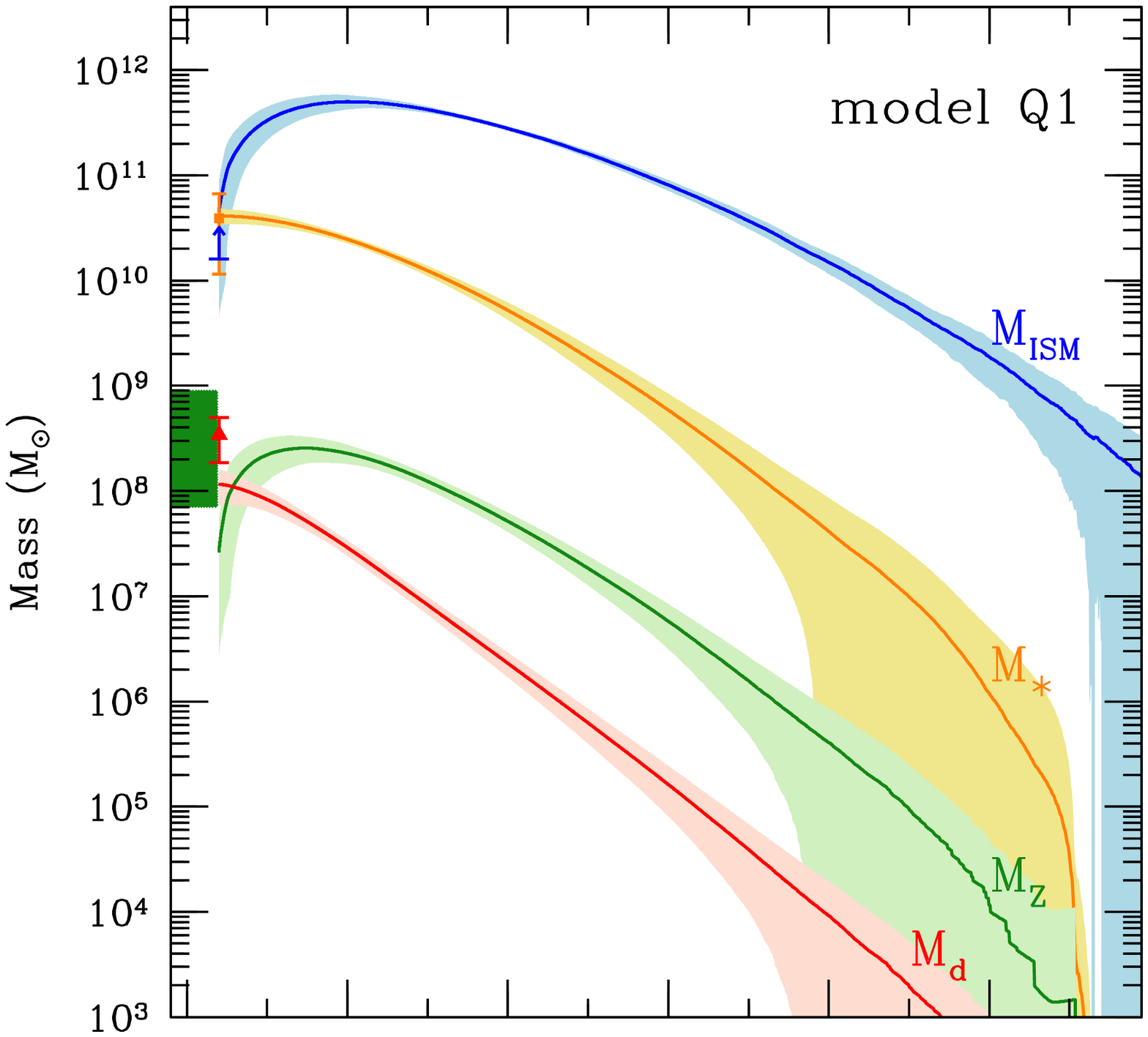}
    \includegraphics{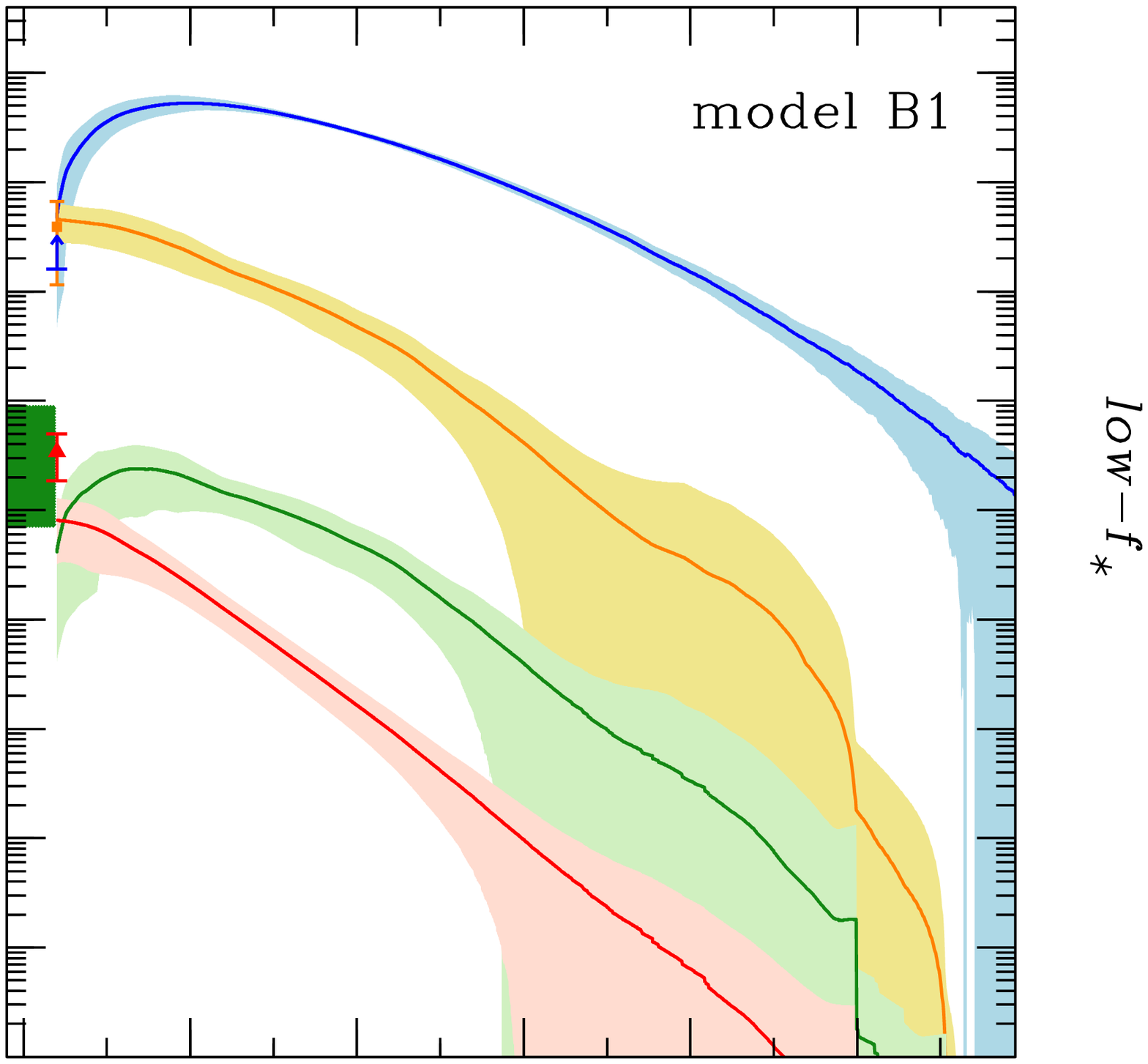}
    \includegraphics{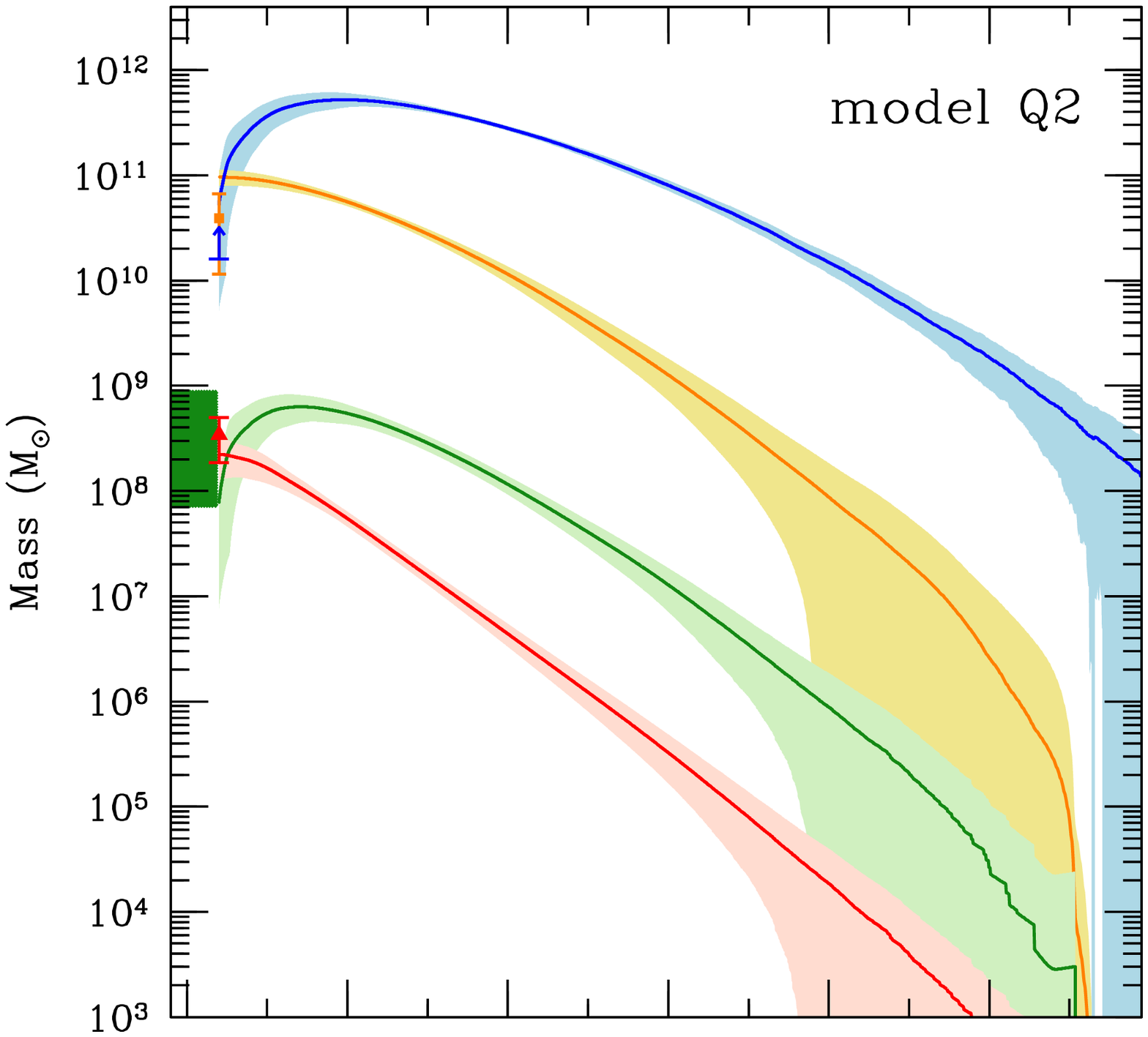}
    \includegraphics{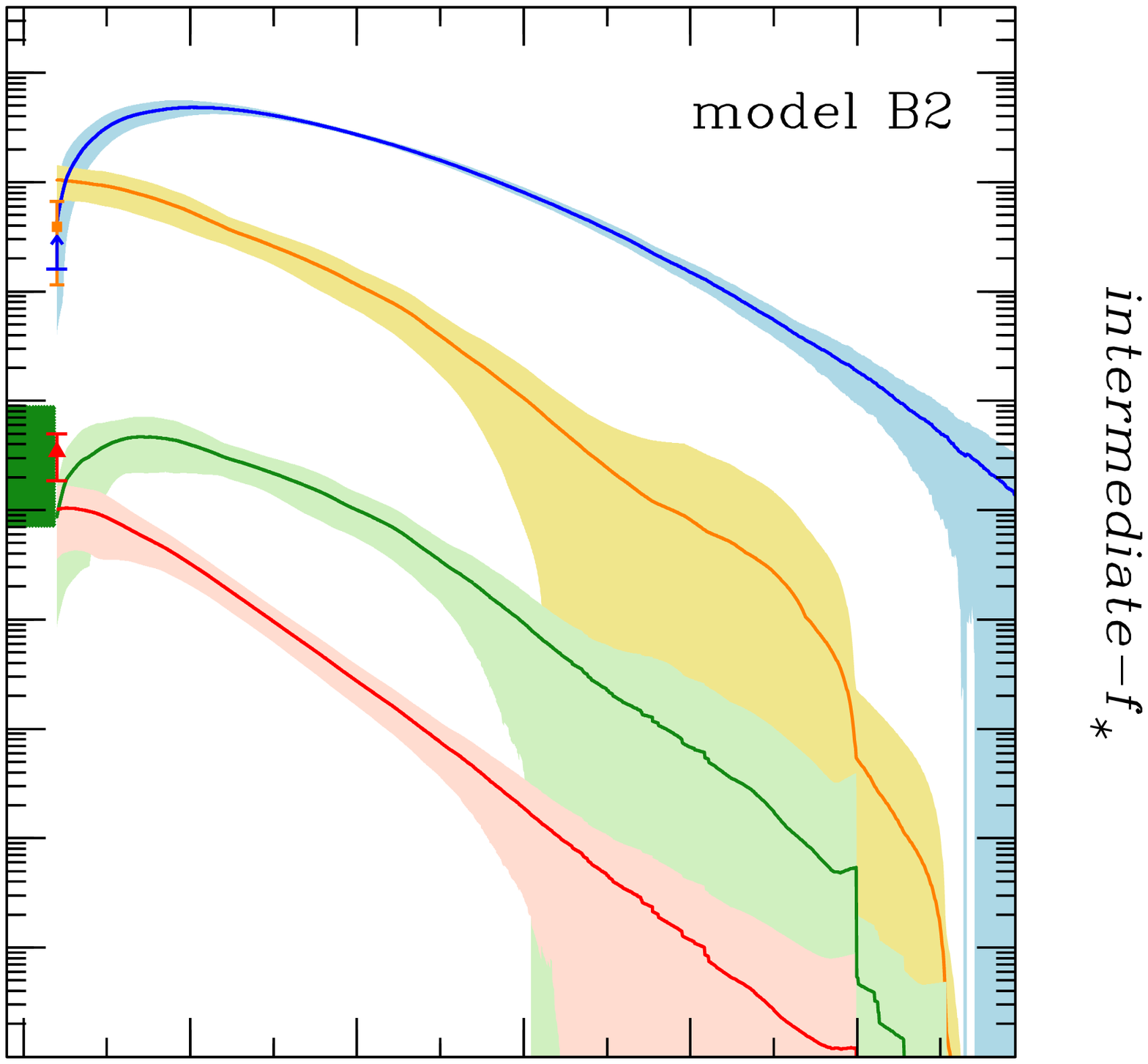}
    \includegraphics{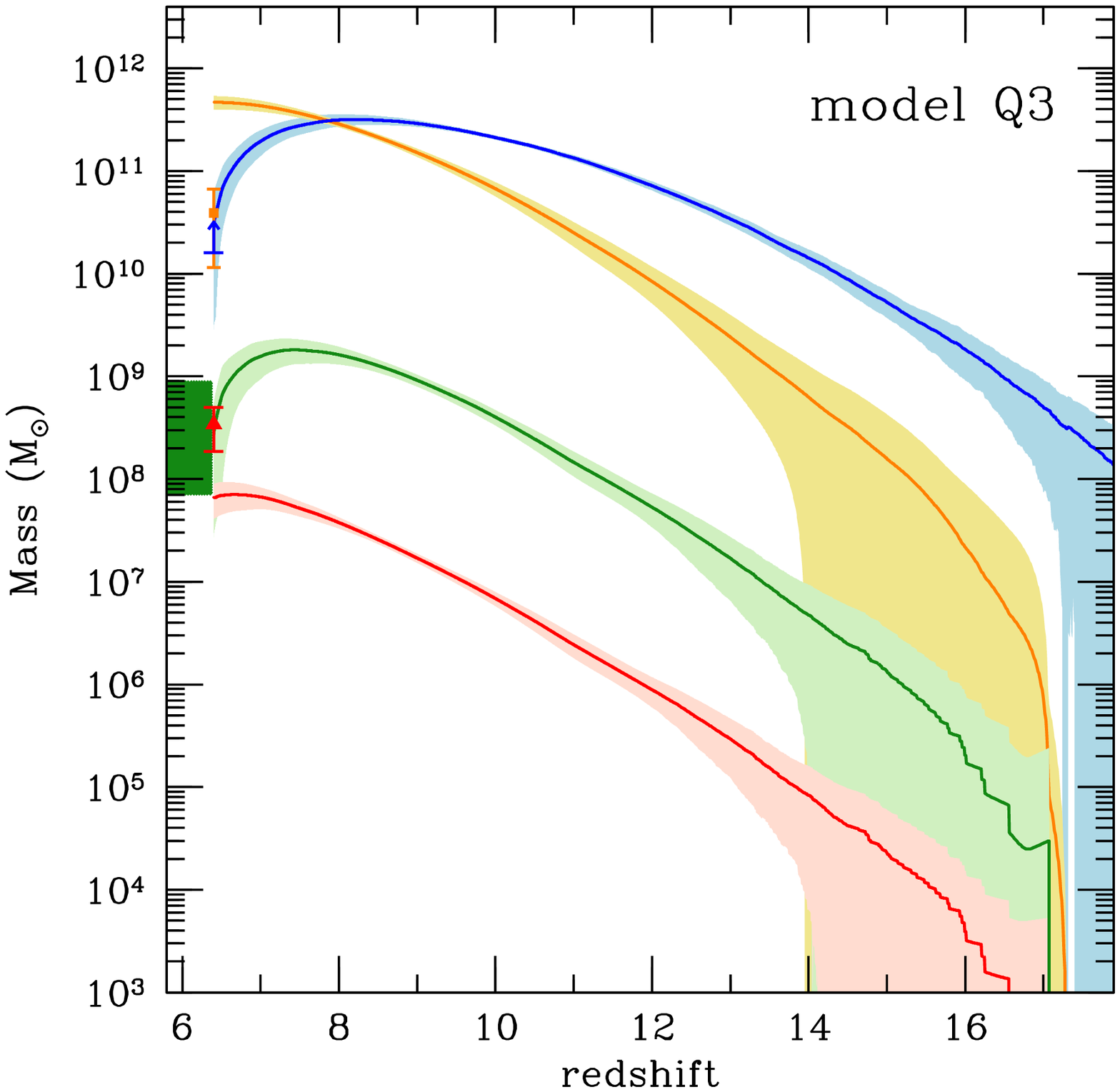}
    \includegraphics{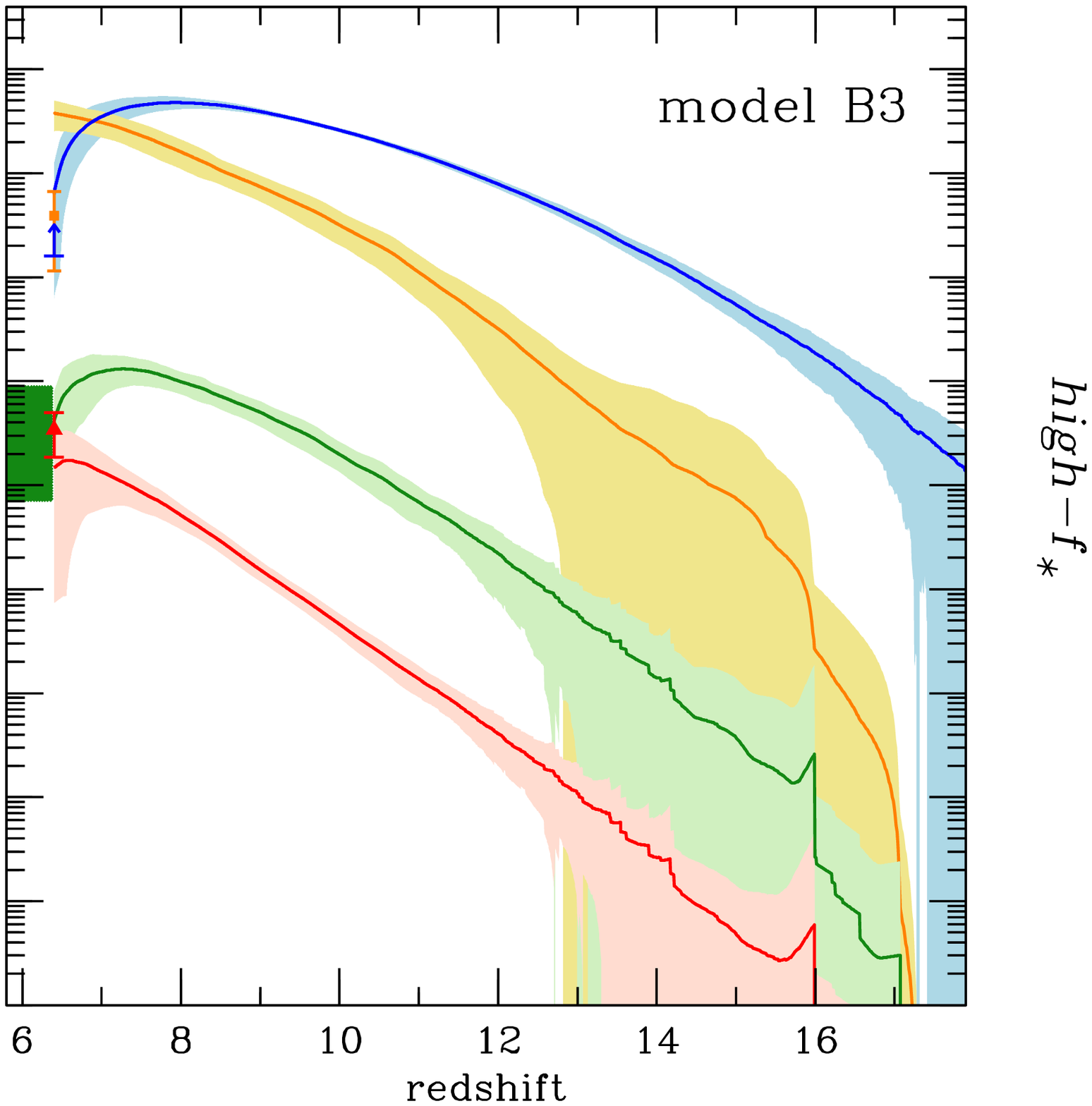}
    \vspace{20cm}
  \end{center}
  \caption{\small
    The ISM chemical evolution for the quasar J1148 host galaxy in different 
    models.
    The lines show the redshift evolution of the average mass of gas, 
    stars, metals and dust.
    {\it Upper panels}: low-$f_\ast$ models Q1 (left) and B1 (right);
    {\it middle panels}: intermediate-$f_\ast$ models Q2 (left) and B2 (right);
    {\it lower panels}: high-$f_\ast$ models Q3 (left) and B3 (right). 
    In all panels solid lines represent the
    average over 50 merger tree realizations and the shaded regions are the 
    $1\sigma$ dispersion.
    The arrow is the lower limit to the gas mass inferred by observations. 
    The circle indicates the final stellar mass with error bars accounting 
    for the uncertainties on the estimates of the dynamical mass.
    The triangle shows the expected average dust mass with 
    errorbars accounting for the range of values given in the literature 
    (see Table \ref{table:dustmass}). 
    Finally, the filled rectangle represents the estimated metallicity (see 
    text).} 
  \label{fig:chemevo}
\end{figure*}

The evolution of the total mass of gas, stars, metals and dust (obtained, at
each redshift, summing over all existing progenitors) for the
host galaxy of J1148 predicted by the six models is shown in Fig. \ref{fig:chemevo}. 
We can immediately see that, independently of the assumed SFH, the mass of 
stars monotonically increases at decreasing redshift. On the contrary,
the mass of gas and metals reach a maximum value at redshift $z\sim 8$, and 
then rapidly decrease as a consequence of the strong AGN feedback (see also 
section \ref{sec:sfr}).
It is interesting to observe that, even for the evolution of the physical 
properties, B-models (right panels) show a higher dispersion among different 
hierarchical merger histories than Q-models (left panels). 
We can furthermore note that, except for the gas mass which is reproduced
by construction, the final properties of the host galaxy strongly 
depend on the assumed SF efficiency, $f_\ast$. 
\newline

\noindent {\it Low-$f_\ast$ models} are the only ones which predict a final 
stellar mass in good agreement with that estimated from the observed
dynamical and molecular gas masses of J1148 ($M_\ast \sim 4\times 10^{10}$ 
M$_\odot$). However, both fail to reproduce the final mass of dust and 
metals, which are under-predicted by a factor of $2-4$. This result implies 
that, independently of the SFH, models which reproduce the observed stellar 
mass are not consistent with the observed chemical properties, at least if an 
ordinary IMF is assumed (see section \ref{sec:chemevo2}). 
\newline

\noindent {\it Intermediate-$f_\ast$ models} predict a final stellar mass
($M_\ast \sim 10^{11}$ $\rm M_\odot$) larger than observed 
but still inconsistent with the value expected from the local
$M_{\rm BH}-M_\ast$ relation (see Fig. \ref{fig:bhrel}).
As a consequence of the larger stellar mass, the total mass of metals produced
increases and it is in (marginal) agreement with the observed value.
The final dust mass is only reproduced by model Q2.   
\newline

\noindent {\it High-$f_\ast$ models} predict a final stellar mass in agreement 
with what expected from the local $M_{BH}-M_\ast$ relation 
[$M_\ast \sim (4-5)\times 10^{11}$ M$_\odot$] and reproduce the observed mass of 
metals. Note that,
despite the large amount of metals available to accrete onto dust grains, the 
quiescent SFH model, Q3, does not reproduce the observed mass of dust.
This result, which is opposite to what found for the intermediate-$f_\ast$ case,
implies that an increase in the total stellar mass by a factor of $4-5$ does not
necessarily produce a similar enhancement in the total amount of dust.
The origin of this behaviour will be discussed in the next section.

\subsubsection{Evolution of dust}\label{sec:dustresults}
\begin{figure}
  \centering  
  \includegraphics[width=8.8cm]{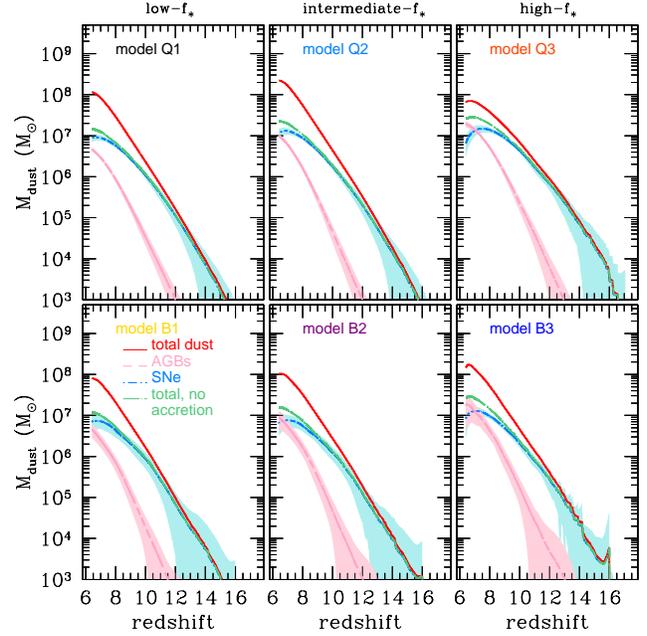}
  \caption{\small
    The evolution of individual dust components as a function of redshift for 
    the {\it low-$f_\ast$} Q1, B1 (left upper and lower panels, respectively), 
    {\it intermediate-$f_\ast$} Q2, B2 (middle upper and lower panels, 
    respectively) and {\it high-$f_\ast$} Q3, B3 (right upper and lower panels, 
    respectively) models. 
    In all panels, the total dust mass (red lines) is compared with 
    {\it (i)} the dust mass computed without the \emph{MC-grown} 
    dust contribution (green lines), {\it (ii)} the \emph{SN dust} (cyan lines)
    and {\it (iii)} the \emph{AGB dust} components (pink lines). All lines
are computed averaging over 50 independent merger trees and the shaded 
    regions represent the $1\sigma$ dispersions for SN and AGB-dust components.} 
  \label{fig:dustcomp}
\end{figure}

In this section we investigate the evolution of the three components that
contribute to the total dust content:
{\it (i)} the mass of dust grown in molecular clouds from seed grains 
produced by stellar sources ({\it MC-grown dust}); 
{\it (ii)} the mass of dust produced by AGB stars ({\it AGB dust}); 
{\it (iii)} the mass of dust formed in SN ejecta ({\it SN dust}).
The evolution of each component along with the total mass of dust (solid red
line) are shown in Fig.~\ref{fig:dustcomp} for all the models. 

It is clear from the figure that dust grown in MCs from seed 
grains produced by stellar sources is the dominant dust 
component at redshift $< 10$. Indeed, the mass of dust contributed by 
stellar sources (green dot-dashed lines) is always sub-dominant at redshift 
$z< 8-10$, and thus cannot reproduce the observed dust mass. 

The evolution of the different components
depends on the SFHs (bursted/quiescent) and star formation efficiencies 
(low-/intermediate-/high-$f_\ast$ models). 
Focusing on stellar dust only, it is interesting to note that a factor 
$\sim 10$ increase in the final mass of stars (going from low- to high-$f_\ast$ models) 
does not imply a similar increase in the final dust mass: the average dust mass 
produced by stellar sources (green lines) increases less than a factor 3 and it 
is almost independent of the SFH. Conversely, the relative importance of AGB 
stars and SN depends both on the SFH and on the star formation efficiency.
In model B2 and high-$f_\ast$ models, AGB stars dominate the stellar dust 
production in the last [150 - 200] Myr of evolution, 
after the peak of star formation and when the effects of SN 
and AGN feedback lead to a strong reduction of the available gas mass. 
Interestingly, the mass of dust produced by AGB stars is found to increase
in models with a higher star formation efficiency while a similar effect is
not observed for SN dust.
Indeed, at redshift $z = 6.4$ the mass of SN dust in high-$f_\ast$ models is 
comparable or smaller than in low- and intermediate-$f_\ast$ models.
This effect can be clarified by the following very simple analytic argument:
assuming that SN are the only dust producers, we can integrate eq.~\ref{eq:dust} 
neglecting the last two terms and applying the instantaneous 
recycling approximation to obtain the "effective" dust yield,

\begin{equation}
y^{\rm eff}_{\rm d, SN} = \frac{M_{\rm dust, SN}}{M_{\ast}} \sim y_{\rm d, SN} - Z_{\rm d}  ( 1+\epsilon_d M_{swept} 
f_{\rm SN} {\cal N}_{\rm SN})
\label{eq:ydeff}
\end{equation}
\noindent
where $y_{\rm d, SN}$ is the IMF-integrated yield of all stars with masses 
$\ge 8 M_\odot$, ${\cal N}_{\rm SN}$ is the number of SN formed per unit mass of 
stars and the other symbols have been introduced in section \ref{sec:dustevo}. 
Thus, as the dust-to-gas ratio, $Z_{\rm d}$, grows in time, the effective dust yield 
decreases for the combined effect of astration and destruction due to SN 
shocks. The larger star formation efficiencies in high-$f_\ast$ models lead to 
higher dust-to-gas ratios (rapid early dust enrichment and higher gas 
consumption) and are both characterized by lower effective dust yields with 
respect to models with lower $f_\ast$. 
For a typical dust yield adopted in the present study 
(see Fig.~\ref{fig:yields}), it is found that $y^{\rm eff}_{\rm d, SN} \sim 0$ when 
$Z_{\rm d, cr} \sim 10^{-4}$. 
Thus, an order-of-magnitude estimate of the maximum mass of dust produced by 
SN can be obtained as $Z_{d, \rm cr} \times M_{\rm gas, max} \sim 5 \times 10^{7} 
M_{\odot}$ where we have assumed the maximum mass of gas shown in
Fig.~\ref{fig:chemevo}. 
This upper limit is consistent with the evolution shown in 
Fig.~\ref{fig:dustcomp}. 

Since AGB stars evolve on longer timescales, their net contribution is less 
affected by astration/destruction and the final mass of dust they produce 
increases with the increasing stellar mass (star formation efficiency). 
However, this can only partly compensate for the reduced 
SN contribution, resulting in total stellar dust masses which vary less than 
30\% among different models and never exceed $3 \times 10^7 M_\odot$. 
We conclude that in all models MC-grown dust is required in order to 
reproduce the observed dust mass. It was already suggested, although not 
confirmed, that dust grown in MCs could give an important contribution to the 
observed dust mass in J1148, given the large molecular gas mass observed in this
systems (Draine 2009). We find that accretion of dust in MCs can easily 
increase the stellar dust mass by a factor $\sim 10$ or more, reaching final 
values up to $\sim 2\times 10^8 M_\odot$, the minimum required to match the 
observations. 

\begin{figure}
  \centering
  \includegraphics[width=8.8cm, height=7cm]{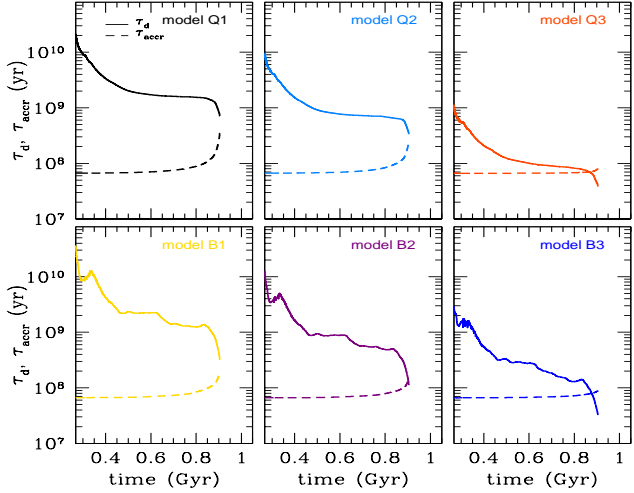}
  \caption{\small
    Time evolution of the accretion (dashed lines) and destruction (solid lines)
    timescales for quiescent models, Q1, Q2, Q3 (upper panels, from left to 
    right) and bursted models, B1, B2, B3 (lower panels, from left to right).} 
  \label{fig:timescales}
\end{figure}

Whether or not a model can reach this limit depends on the 
relative importance of the timescales for dust destruction and accretion, 
$\tau_{d}$ and $\tau_{acc}$.
In Fig.~\ref{fig:timescales} we show these two timescales as a function of 
time for all the models discussed so far: 
in all cases is $\tau_{acc} << \tau_{d}$, with the exception of the latest 
stages of the evolution when, despite the decline of the supernova rate, gas 
and metal ejection due to AGN feedback conspire to decrease the destruction 
timescale (less supernovae are required to shock the remaining gas) and increase
the accretion timescale (less metals are available to be accreted onto dust 
grains, hence $f_{\rm dep} \rightarrow 1$ in eq.~\ref{eq:tauacc}). 

As expected, $\tau_d$ is progressively shorter in models where the increase of
the star formation efficiency leads to larger supernova rates. Thus, in models
Q1, B1, and Q2 the final dust mass is controlled by accretion (limited by the
available mass of metals), in models B2, Q3 and B3 it is limited by
destruction.\\ 

\subsubsection{The contribution of Population III stars to metal and dust enrichment}\label{sec:pop3}

In this section we analyze the contribution of Pop III stars to metal and dust
enrichment of the ISM of the QSO host galaxy. We discuss the results obtained in
model Q1 since other models provide very similar conclusions.

\begin{figure}
  \centering
  \includegraphics[width=4.4cm]{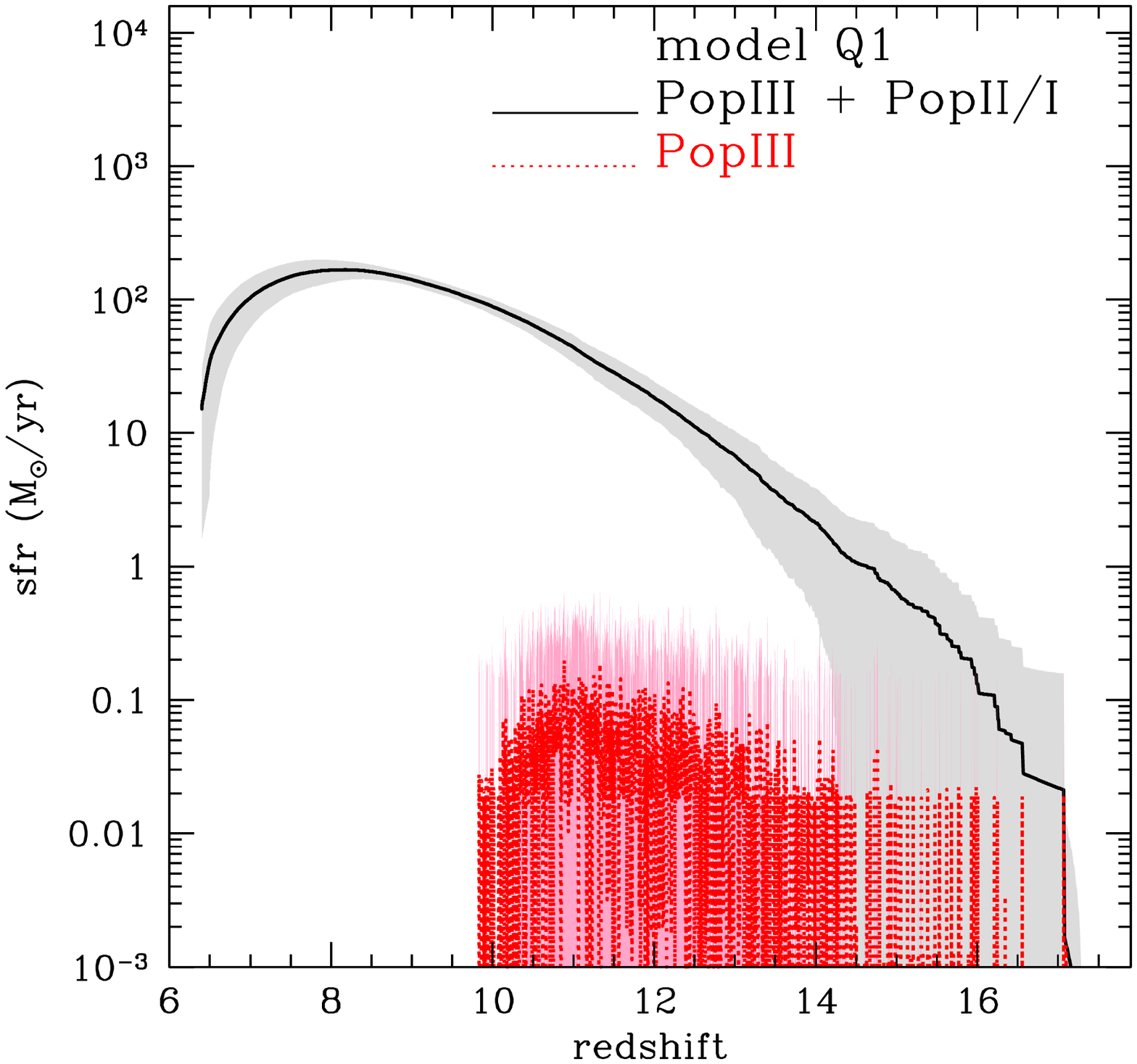}\includegraphics[width=4.4cm]{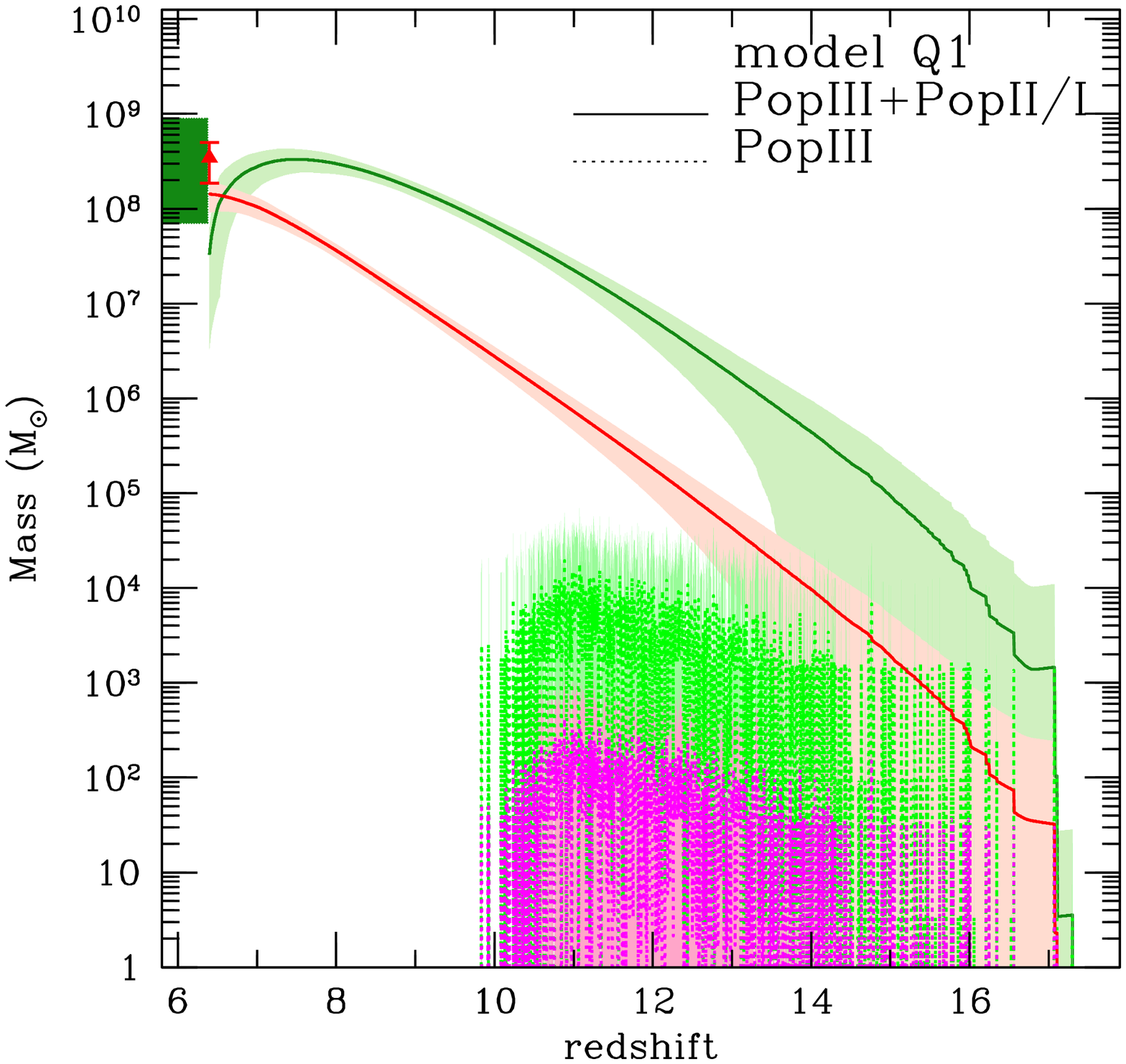}
  \caption{\small
    The contribution of the first stars to chemical enrichment of the host 
    galaxy ISM as predicted by model Q1. 
    Left panel: the SFH of Pop III stars (dotted line) as a function of redshift
    is compared with the total one (solid line).
    Right panel: the  mass of metals (green dotted line) and dust (magenta 
    dotted line) produced by Pop III stars at each redshift is compared to the
    total ones (solid ones).
    The shaded areas represent the corresponding $1\sigma$ dispersion
    of each quantity.} 
  \label{fig:pop3}
\end{figure}

The Pop III star formation rate as a function of redshift is shown
in the left panel of Fig.~\ref{fig:pop3} (red dotted line with $1-\sigma$
dispersion). The solid line represents the total (Pop III and Pop II/I) star 
formation rate (same as in the left upper panel of Fig.~\ref{fig:sfr}).
As it can be seen from the figure, the formation of Pop III stars is dominant
only in the earliest phases of the evolution ($z>16.5$) when the gas metallicity
in progenitor star-forming halos is sub-critical ($Z<Z_{\rm cr}$, 
see section \ref{sec:model_summary}). In these halos, the gas 
is promptly enriched by PISNe to metallicities $Z>Z_{\rm cr}$, triggering the formation of
lower mass stars; the formation of Pop III stars is still active 
down to redshift $\sim 9.8$. However, such starbursts typically occur in a small
number of halos at each redshift and the resulting Pop~III star formation rate 
is orders of magnitude less efficient with respect to the total one. 

The right panel of Fig.~\ref{fig:pop3} shows the mass of metals and dust contributed by Pop III
stars at each redshift (green and magenta dotted lines, respectively) with shaded areas 
indicating the corresponding $1\sigma$ dispersion. PISNe provide up to 100\% of the
total mass of metals and dust at $z>16$ but their contribution rapidly drops with time
and at the final redshift less then [5 - 10]\% of metal/dust mass is released by 
Pop~III stars.

Thus, given their negligible contribution, we conclude that the choice of
the characteristic mass for these stars does not influence the chemical 
evolution of the QSO host galaxy. This result is not surprising since,
as already emphasized by Schneider, 
Venkatesan \& Ferrara (2004), QSOs form in overdense regions of the Universe 
where the Pop III/Pop II transition occurs promptly and the enrichment is 
dominated by Pop II/I stars. 

\subsection{Chemical evolution with a Top-heavy IMF}\label{sec:chemevo2}
We have shown that low-$f_\ast$ models, which reproduce the final 
stellar mass inferred from observations of J1148, fail to reproduce 
both the observed dust and metal masses (see Fig.~\ref{fig:chemevo}, upper 
panels). 
Depending on the adopted SFH, a factor of $\sim 3$ (Q2 model, Fig.~\ref{fig:chemevo}, middle left panel) to 
$\sim 10$ (B3 model, Fig.~\ref{fig:chemevo}, lower right panel) higher final 
stellar mass is required to reproduce the observed chemical properties of 
J1148, if a Larson IMF with $m_{ch}=0.35$~M$_\odot$ is adopted.

\begin{figure}
  \centering
  \includegraphics[width=9.0cm]{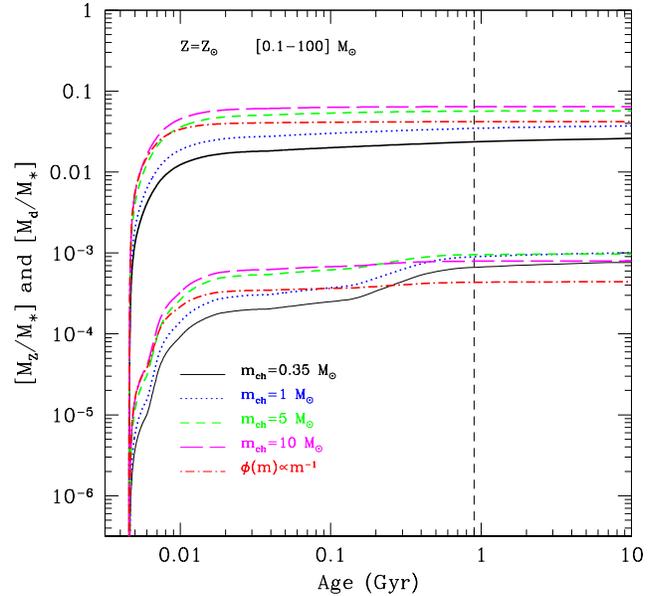}
  \caption{Evolution of the metal (upper lines) and dust (lower lines) yield 
    as a function of stellar age. All stars are assumed to form in a single 
    burst at $t=0$ with solar metallicity with a Larson IMF with 
    characteristic masses $m_{ch} = 0.35, 1, 5, 10 \rm M_\odot$ and with a 
    {\it flat} IMF (see text). 
    The IMFs are normalized to unity in the mass range $[0.1-100] \rm M_\odot$. 
    The vertical line indicates the age of Universe at redshift $z=6.4$.}
  \label{fig:yields}
\end{figure}

In this section, we investigate how these results depend on the stellar IMF.
Fig.~\ref{fig:yields} shows the evolution of total metal  
(upper curves) and dust (lower curves) yields as a function of stellar age,
assuming that all stars are formed in a single starburst at $t=0$ with solar
metallicity (see eq.~1 in V09). 
Yields are computed for a Larson IMF with different characteristic masses, 
$m_{ch}= 0.35, 1, 5, 10  \rm M_\odot$, and for a {\it flat} IMF, 
$\phi(m)\propto m^{-1}$. 
This latter IMF has been proposed to reconcile predictions of semi-analytical
models with the observed number counts of submillimeter galaxies at high
redshift (e.g. Baugh et al. 2005, but see also Hayward et al. 2010).
IMFs are all normalized to unity in the mass range $[0.1-100] \rm M_\odot$ and
the vertical line indicates the age of Universe at redshift $z=6.4$.

We find that a Larson IMF with $m_{ch}=5 \rm M_\odot$ provides the most 
favourable conditions for larger amount of dust and metals to be 
contemporarily produced by a single stellar population: a factor of $\sim 3$ 
larger metal yield is produced when $m_{ch}$ is increased from our standard 
value $0.35 \rm M_\odot$ to $m_{ch} = (5-10) \rm M_\odot$. 
The highest amount of dust per unit stellar mass formed
($\sim 10^{-3}$) at $z = 6.4$ is obtained when a Larson IMF with 
$m_{ch}=5 \rm M_\odot$ is adopted, which still includes a non negligible 
contribution from AGB stars, comparable to that of SNe (see also V09). 

\begin{figure*}
  \centering
  \includegraphics[width=7.0cm]{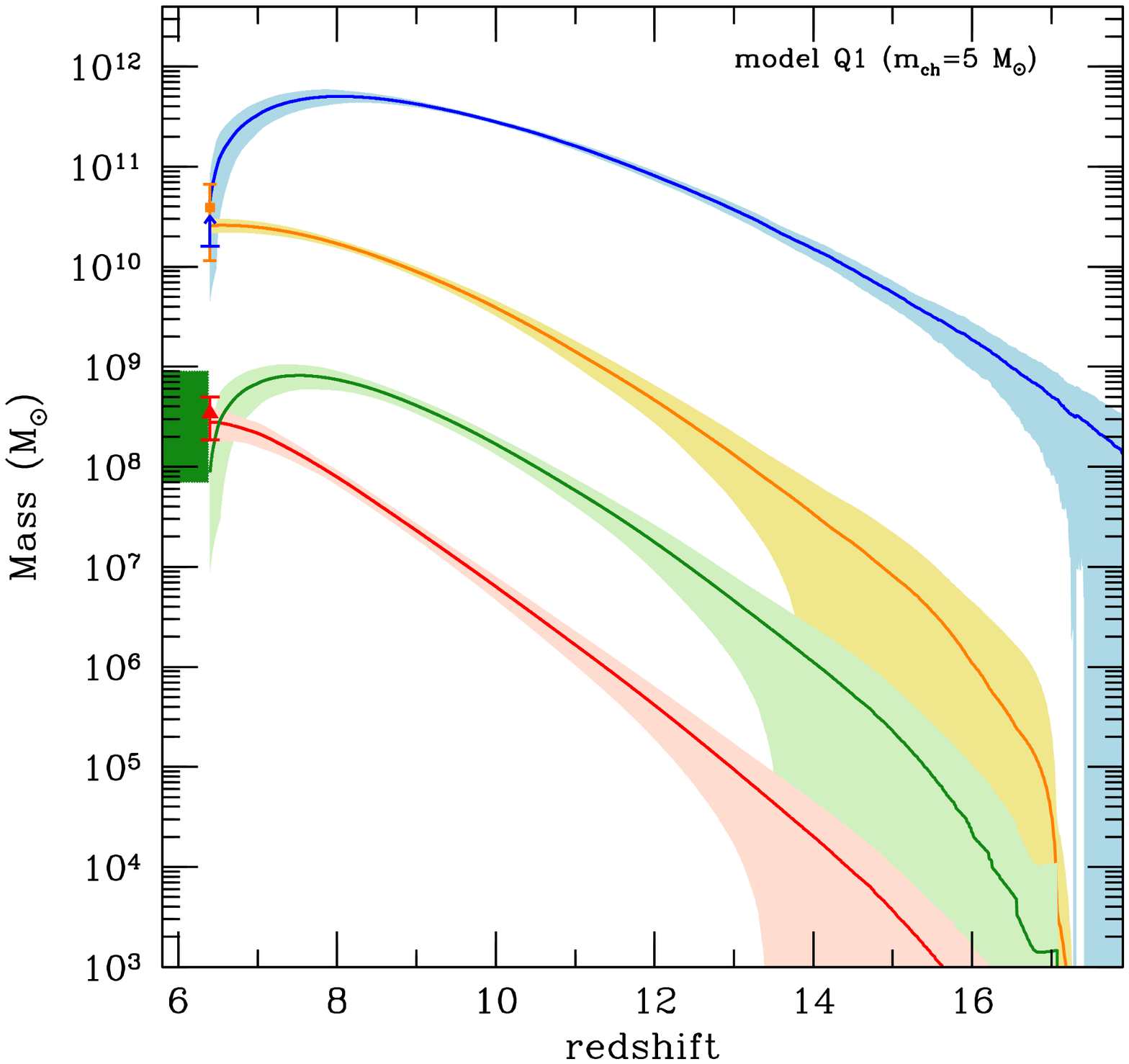}\includegraphics[width=7.0cm]{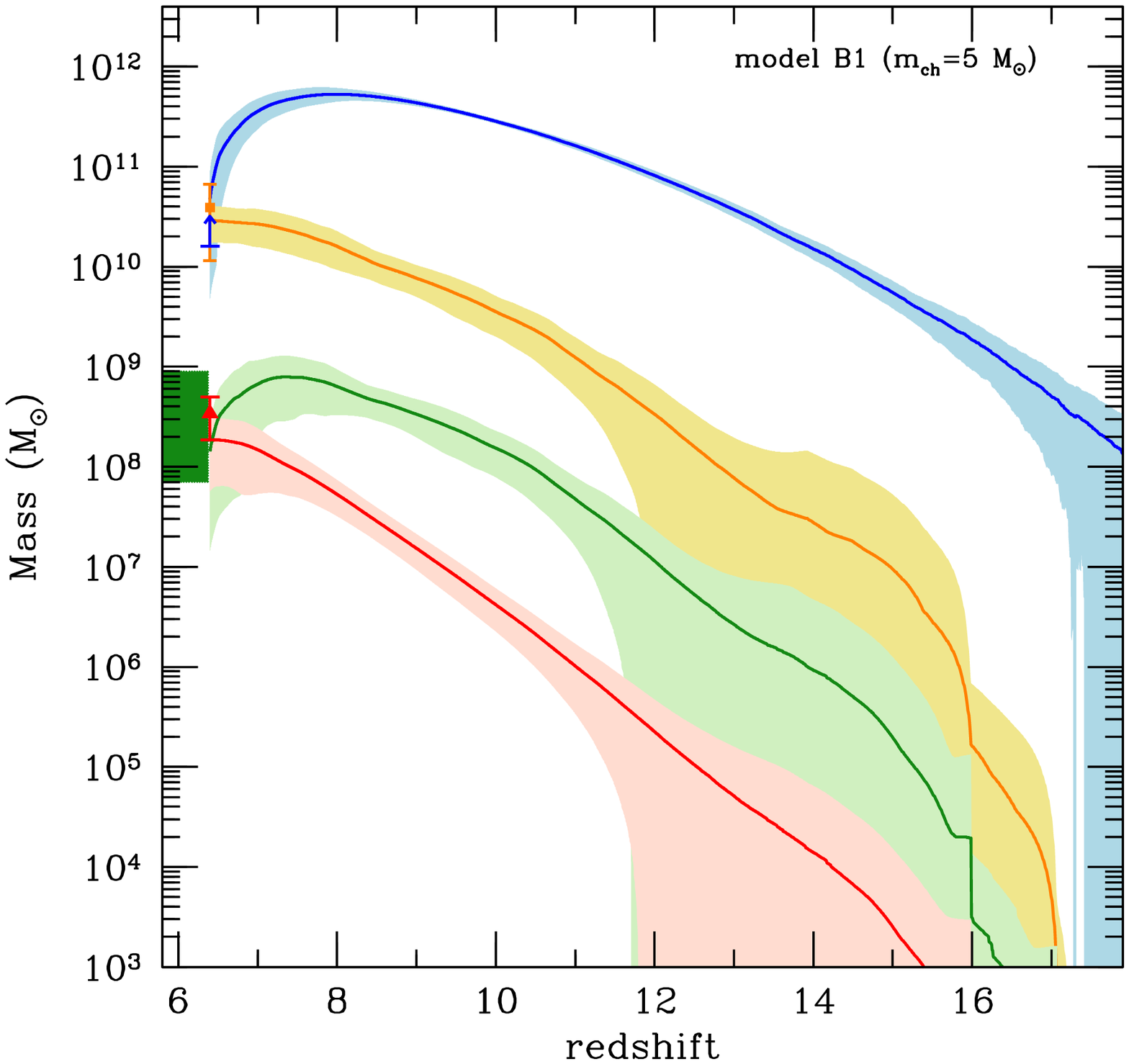}
  \caption{\small
    The same as in Fig~\ref{fig:chemevo} but for {\it low-$f_\ast$} models
    with a Larson IMF with $m_{ch}=5 \rm M_\odot$.
    } 
  \label{fig:lowchem_mch5}
\end{figure*}

Fig.~\ref{fig:lowchem_mch5} shows the resulting chemical properties for 
low-$f_\ast$ models when a Larson IMF with $m_{ch} = 5 \rm M_\odot$ is adopted. 
With this {\it top-heavy} IMF, a factor of $\sim 2-3$ larger dust and metal 
masses at the final redshift $z=6.4$ are produced compared to the same models 
with a {\it standard} IMF with $m_{ch}=0.35 \rm M_\odot$ (Fig.~\ref{fig:chemevo}, 
upper panels).
A higher characteristic mass, $m_{ch}$, implies a larger number of intermediate- 
and high-mass stars. In fact, supernova rates comparable to those found in 
intermediate-$f_\ast$ models, which are characterized by a higher star 
formation efficiency, are produced.

\begin{figure}
  \centering  
  \includegraphics[width=4.4cm]{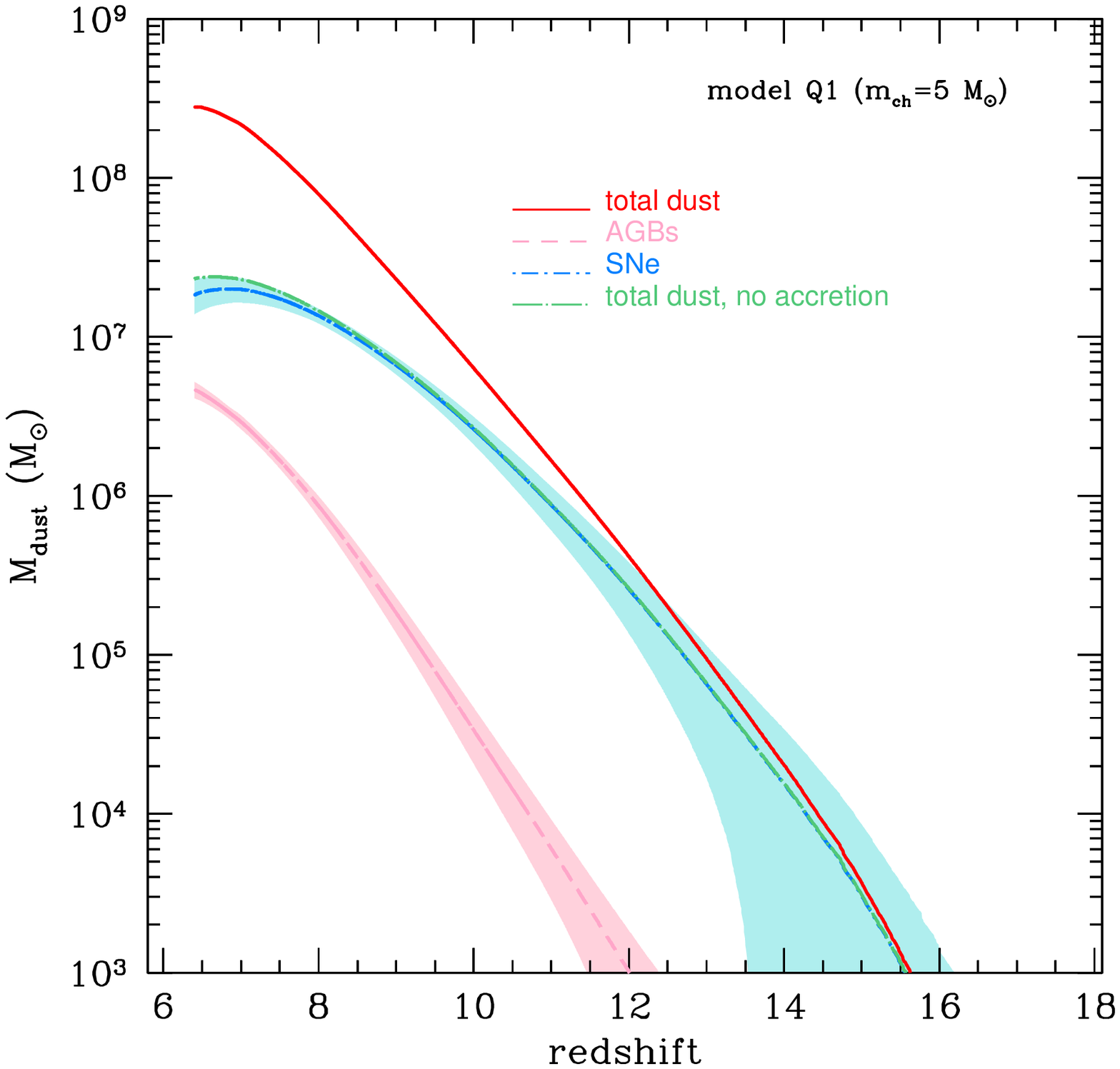}\includegraphics[width=4.4cm]{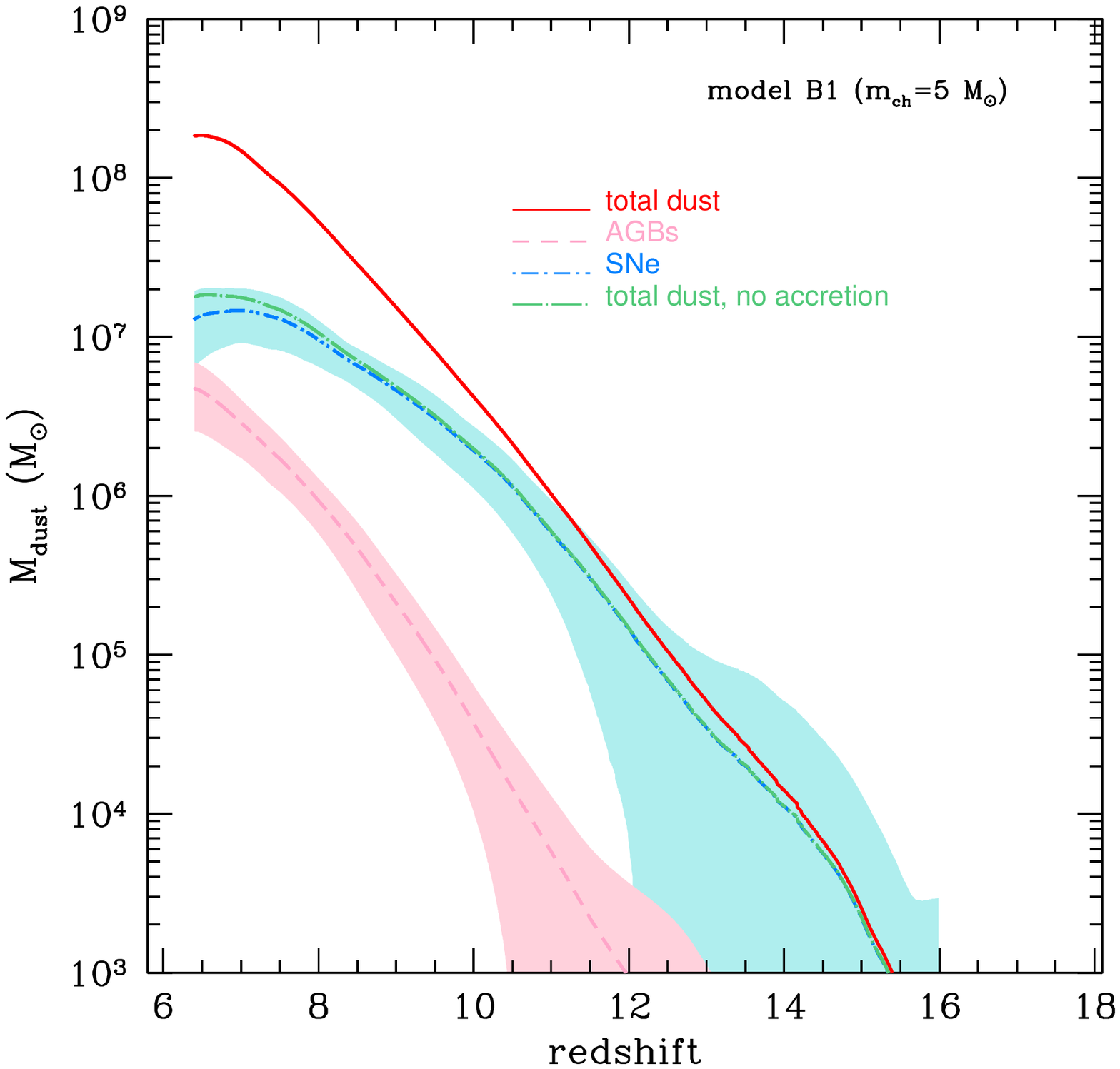}
  \caption{\small
    The same as in Fig.~\ref{fig:dustcomp} but for low-$f_\ast$ models with 
    a top-heavy IMF (Larson IMF with $m_{ch}=5 \rm M_\odot$).} 
  \label{fig:dustcomp_mch5}
\end{figure}

Since SNe have the dual role of both producing and destroying dust, the 
higher dust yield leads only to a moderate increase in the total mass of 
stellar dust. In Fig.~\ref{fig:dustcomp_mch5}, we show
the time evolution of individual dust components; the change in the IMF
causes SNe to dominate stellar dust production at all redshifts. Yet, the
total mass of stellar dust is still $\sim 2 \times 10^7 M_{\odot}$, consistent
with the argument\footnote{It is important to note that although the SN dust 
yield $y_{d, SN}$ is higher when $m_{\rm ch}=5 \rm M_\odot$, the number of SN 
formed per unit mass of stars, ${\cal N}_{\rm SN}$, is also higher
resulting in a $Z_{\rm d, cr}$ comparable to the value found for a standard IMF.}
given by eq.~\ref{eq:ydeff}. The larger mass of metals available allows
$\sim [2 - 3] \times 10^8 M_{\odot}$ mass of dust to grow in MCs; by the end
of the simulation, an equilibrium in reached where $\tau_d \sim \tau_{acc}$; 
the onset of this equilibrium limits the mass of dust. 
  
Thus, the observed chemical properties of J1148 seem to require either models 
with a factor $3 - 10$ larger stellar mass, which would shift this high redshift
QSO close or onto the observed local stellar bulge-black hole mass relation,
or models with a non-standard IMF, in which stars form according to a Larson IMF
with a high characteristic mass, $m_{ch}=5 \rm M_\odot$. 
 
\section{Discussion}
\label{sec:discussion}
Several theoretical models aimed to investigate the formation and evolution of 
dust in high redshift ($z > 5$) galaxies and quasars have been 
developed (e.g. Morgan \& Edmunds 2003; Hirashita \& Ferrara 2002; Dwek et al 
2007; V09; Gall et al. 2010; Dwek \& Cherchneff 2010; Gall et al. 2011; Mattson
2011). Models which reproduce the observed dust mass in J1148 using SNe as the 
primary sources require high dust yields ($0.3 \rm M_\odot - 1 M_\odot$), special 
SFHs (all stars formed in a starbursts with age 0.3 - 100 Myr), top-heavy 
IMF and/or moderate dust destruction by interstellar shocks (Dwek, Galliano
\& Jones 2007; Dwek \& Cherchneff 2011; Gall et al. 2010, 2011). 
The high dust yields required by these theoretical calculations are not 
supported by observations of local SNe (Wooden et al. 1993;
Elmhamdi et al. 2003; Pozzo et al. 2004; Ercolano et al. 2007; 
Meikle et al. 2007; Kotak et al. 2009) and SN remnants. 
Among the latter class, the best studied source is Cas A which has been 
recently observed with Spitzer (Rho et al. 2008) and Herschel (Barlow et al. 
2010), yielding an estimated dust mass in the range 
$2 \times 10^{-2} {\rm M_{\odot}} < M_{\rm dust} < 7.5 \times 10^{-2}{\rm M_{\odot}}$
(see however Dunne et al. 2009, who proposed the presence of 
$>0.1 \rm M_{\odot}$ of dust in Cas A based on submm polarization observations 
with SCUBA).

Additional dust sources have been also proposed. In V09 we pointed out that 
AGB stars give an important contribution to the observed dust mass at $z>6$
using a SFH for the host galaxy of J1148 as predicted by numerical simulations
(Li et al. 2007).  This finding has been recently confirmed by 
Dwek \& Cherchneff (2011) and by Pipino et al. (2011); the latter 
also show that, on galactic scales, dust produced by the AGN itself 
- a mechanism known as smoking QSOs (Elvis et al. 2002) - 
is negligible when compared to that of stellar sources. 
Quantitative comparisons among various studies are difficult due to
different prescriptions for computing dust formation
and evolution (SFH, IMF, dust and metal yields, dust destruction efficiencies). 
However, all the investigations published so far underline the important role
played by the adopted SFH and IMF.
 
In this work we have shown that MC-grown dust is the 
dominant component even at early cosmic epochs, $z > 6$, while stellar 
sources, SNe and AGB stars, can not account for the total amount of dust 
observed in J1148. Their relative contribution is determined by the assumed 
SFH and IMF.
This result differs from what was inferred by the simple toy-model presented
in V09 for two main reasons: the peculiar input SFH and
the close-box approximation made in V09.
Specifically, the redshift of the last major burst of star formation 
and its intensity appear to provide the optimal conditions for dust 
production by stellar sources. In turn, our new results show the 
importance of a statistical analysis, where a sufficiently large 
sample of SFHs (merger tree realizations) can be investigated. 
In addition to that, galactic outflows determine the average
ISM dust-to-gas ratio and have an indirect effect on the dust destruction 
timescales. 

It has been shown in V09 that higher dust masses can be produced 
decoupling the contribution of AGB stars from that of SNe. 
Indeed, we find that the largest stellar dust masses are found in some 
realizations of model B3 where the epoch 
and amplitude of the last major burst maximize dust production by AGB stars. 
Yet, when the average over many independent realizations is considered, the 
effect is partly washed out (see Appendix A).

Exploring different possible star formation histories for the QSO host galaxy,
we find that assuming a Larson IMF with $m_{ch} = 0.35 \rm M_\odot$, 
the observed chemical properties (dust mass and gas metallicity) of J1148 
seem to require a final stellar mass which is a factor $3 - 10$ 
higher than what inferred by observations of the dynamical and molecular gas 
masses. It is well known that observations of the dynamics and spatial
distribution of molecular gas in J1148 (Walter et al. 2003), together with
those which apply to a larger sample of QSOs with $5 < z < 6$ (Wang et al.
2010) seem to suggest a BH/stellar mass ratio which is 10-20 times higher than
the present-day value, implying that the central BHs in these systems may
assemble before their stellar bulge.
Observations of dynamical and molecular gas are limited to
the central 2.5 kpc of the host galaxy; a factor 10 increase in the
estimated stellar mass could be made consistent with the observations
simply assuming that the stellar bulge extends out to 25 kpc for realistic
radial density distributions. More importantly, {\it the required stellar
mass - based on chemical properties of the host galaxy - would
shift the data point of J1148 on the BH-stellar mass relation much
closer to the observed correlation in the present-day galaxies,
at least within the observed scatter.}

Alternatively, all the observed properties of J1148 and its host galaxy 
can be reproduced if Pop II stars are assumed to form according to a 
Larson IMF with $m_{ch} = 5 \rm M_{\odot}$, independently of the assumed SFH.
Since SNe both form and destroy dust grains, the larger dust yield due to
higher supernova rate leads only to a mild increase in the total dust mass
produced by stellar sources. Yet, the larger metal yields lead to larger 
masses of gas-phase metals, which allow seed dust grains to efficiently 
grow in MCs. 
Indeed, physical conditions present in the ISM of QSO host galaxies at high-$z$ 
may favour a shift in the fragmentation mass scale of star forming clouds, 
hence in the characteristic mass of the stellar IMF.
Schneider \& Omukai (2010) have studied the response of the thermal Jeans mass
at cloud fragmentation to the joint effect of metal-line cooling, dust-cooling
and the cosmic microwave background (CMB) in the ISM of high-redshift galaxies. 
They show that when $z \ge 5$ and the metallicity is $Z \ge 10^{-3} Z_\odot$, 
heating of dust grains by the CMB shifts the Jeans scale to larger masses. 
If no dust grains are present in star forming cloud, the effect is even more 
dramatic and only massive and very massive stars are predicted to form (Smith 
et al. 2009; Jappsen et al. 2009).
In addition to the effect of the CMB, MCs in circumnuclear starburst regions 
may be characterized by gas temperatures $\sim 100$~K
and densities of a few $ 10^5$~cm$^{-3}$; under these conditions
numerical simulations predict that the resulting mass spectrum of gravitational
condensations (a proxy to the IMF) is top-heavy, compared to that in the
solar neighbourhood (Klessen, Spaans \& Jappsen 2007).

In the present study (as in V09)
SN dust yields are obtained using one of the models proposed by
Bianchi \& Schneider (2007) where only 7\% of the dust formed in 
SN ejecta survive the passage of the reverse shock. 
According to this model, $\sim [0.5 - 5]\times 
10^{-2})$ M$_\odot$ of dust per SN is injected in the ISM 
(depending on the progenitor star mass and metallicity), 
in good agreement with dust observed in young SN remnants.
It is important to discuss how a different choice would affect the
results present in previous sections. Indeed, if we adopt an alternative
model proposed by Bianchi \& Schneider (2007), where the fraction of
SN-condensed dust which is injected in the ISM is 20\%, we find that
the final dust masses are similarily shifted upward 
by a factor $\sim 3$, bringing models Q1 and B1 in agreement with the
observational data point. However both models severely under-predict the
mass of metals. Thus, although a higher SN dust yield leads to a larger 
total dust mass (closer to the observed value), a larger final 
stellar mass or a top-heavy IMF are still required to reproduce 
all the observed properties of J1148.

A final remark concerns the adopted prescription for the AGN feedback.
Feruglio et al. (2010) have recently 
detected the broad wings of the CO line in Mrk 231, the closest known QSO, 
which has enabled to trace a giant molecular outflow of about 700 
$\rm M_{\odot}/$yr; the inferred kinetic energy in the molecular outflow is 
$1.2 \, 10^{44}$~erg/s, corresponding to a few percent of the AGN bolometric 
luminosity ($5 \, 10^{45}$~erg/s).  
This is consistent with what we find in our simulated models. 
In fact, the formulation adopted for the BH accretion rate
ensures that BH feedback is active over almost the entire lifetime
of the host halo (since the computed rate is almost always above the 
critical threshold), resulting in an efficient gas ejection at lower 
redshift ($z< 7-8$) when the BH mass and its gas accretion rate are higher. 
This, in turn, is regulated by the choice of the parameters $\alpha$ and 
$\epsilon_{\rm w,AGN}$.  
However, it is important to note that the relative role of starburst and 
AGN activity in regulating/quenching star formation and BH-growth, as well as 
the mechanisms (energy/momentum deposition) driving these feedback processes, 
are still largely debated (see e.g. Bhur et al. 2010; Hopkins 2011; Hopkins, 
Quataert \& Murray 2011 for a different view). 

\section{Summary and conclusions}
Here we have presented a semi-analytical model which allows to 
{\it (i)} model the evolution of a QSO host galaxy and its central supermassive
black hole, predicting a range of possible SFHs and to {\it (ii)} explore the 
relative importance of stellar sources of dust (SN and AGB stars) and of dust 
grown in MCs at these early cosmic epochs. 
\textsc{GAMETE/QSOdust} follows the build-up of the central SMBH through 
mergers and gas accretion, starting from seed black holes at very high 
redshifts.
The corresponding assembly of the host galaxy is followed in a 
self-consistent way and the co-evolution of these two fundamental components is 
controlled by AGN feedback in the form of a galactic-scale wind, assuming that 
0.5\% of the rest mass energy of the gas accreted onto the BH is thermally 
coupled to the ISM.
We have performed 50 random hierarchical merger histories of a 
$10^{13}\rm{M_\odot}$ dark matter halo, which is believed to host a $>10^9 
\rm M_\odot$ SMBH, through a binary Monte Carlo algorithm with mass accretion 
based on the EPS theory. 

Metal and dust enrichment is followed evolving all progenitor galaxies on the 
corresponding stellar lifetimes, considering the contribution from Pop III 
(PISNe) and Pop II stars (AGB and SNe), and the subsequent dust grains 
processing (destruction and growth) in the ISM. Metal and dust yields are
taken from theoretical models for stellar nucleosynthesis and dust nucleation
in SN ejecta and AGB stellar atmospheres.
Different SFHs for the QSO host galaxy have been explored: 
in quiescent models, the efficiency of star formation is independent of the 
galaxy mergers mass ratio, while in bursted models the star formation 
efficiency is enhanced during major mergers among progenitor galaxies.
The free parameters of the models, namely the efficiency of star formation, 
BH accretion and AGN-driven wind have been chosen to reproduce the BH and 
gas mass inferred from observations of J1148 at redshift $z=6.4$.

The main results of this study can be summarized as follows:
\begin{itemize}
  \item the chemical properties of the host galaxies of high-$z$ QSOs 
    are a powerful probe of their past evolution. The dust mass produced
    by stars depends on the shape of the star formation history, which is
    controlled by the evolution of the gas content through mergers and 
    outflows driven by SNe and by the AGN. The total mass of metals is a 
    tracer of the integrated star formation history (total stellar mass) 
    and it is sensitive to the strength and frequencies of metal-rich outflows.

  \item Population III stars give a negligible contribution to chemical enrichment of 
    the ISM. In fact, the large metal and dust yields expected from theoretical
    models of PISN lead to rapid enrichment of the ISM to metallicities $Z \ge 
    Z_{\rm cr} = [10^{-6} - 10^{-4}] Z_\odot$ in progenitor galaxies of the final 
    galaxy host, enabling the formation of Population II/I stars.   

  \item If Population II/I stars form with a {\it standard IMF} and a 
    characteristic stellar mass of $m_{ch}=0.35~ \rm M_\odot$, as in the 
    present-day Universe, the final masses of metals and dust are lower than 
    observed. Depending on the adopted SFH, a final stellar mass which is a 
    factor of $3-10$ larger than the value inferred from observations of the 
    dynamics and spatial distribution of the molecular gas content of J1148 
    (Walter et al. 2003) is required. The required stellar mass would shift the
    position of J1148 onto the locally observed correlation between the SMBH
    and stellar bulge masses, at least within the observed scatter.

  \item If Population II/I stars form with a {\it top-heavy IMF} and a 
    characteristic stellar mass of $m_{ch}=5~ \rm M_\odot$ then the observed 
    chemical properties can be reconciled with the inferred stellar mass. 
    The physical conditions of star forming regions in these high-$z$ systems 
    may in fact suppress fragmentation, due to higher temperatures in MCs and 
    heating of dust grains, favouring the formation of larger stellar masses.

  \item A statistical analysis of the SFHs, with many independent
    hierarchical merger trees of the dark matter halo hosting J1148, 
    has allowed us to conclude that although SNe
    dominate the early dust enrichment, AGB stars contribute at $z < 8 -10$;
    the latter contribution depends on the shape of the SFH and on the 
    adopted IMF for Population II/I stars. 

  \item We find that stellar sources produce a total dust mass which never
   exceeds $\sim 2-3 \times 10^7 M_\odot$; in fact, unless the redshift and
   intensity of the last major burst of star formation allows to efficiently 
   decouple the contribution of SNe and AGB stars, the effective dust yield
   is progressively reduced due to the joint effect of astration and dust 
   destruction in the ISM by SN shocks. This conclusion is independent of
   the adopted stellar IMF but depend on the adopted stellar dust yields.
    
  \item In all the models we have investigated, the final dust mass
   can be reconciled with the observed value, $M_{\rm dust} = [2 - 6]\times 10^8 $
   M$_\odot$ via grain growth in molecular clouds which appear to be important 
   even at these high redshifts, consistent with the large molecular gas mass 
   observed in J1148. 
   The final mass of dust is thus controlled by two fundamental timescales: the
   timescale for grain growth (which become progressively longer due to rapid 
   depletion of metals in dust grains) and the timescale for dust destruction 
   by SN shocks (which depends on the SFH through both gas consumption/ejection
   and the SN rate). 

\end{itemize}

\section*{Acknowledgments}
We thank the anonymous referee for useful comments and suggestions.
We are grateful to A. Ferrara, S. Gallerani, R. Maiolino, A. Marconi, 
and G. Risaliti for profitable discussions and suggestions.
We also acknowledge DAVID members\footnote{http://wiki.arcetri.astro.it/bin/view/DAVID/WebHome} 
for fruitful comments. RV whishes to thank L. Zappacosta for precious help.   

\appendix

\section{Dependence on the starburst frequency}
\begin{figure}
  \centering
  \includegraphics[width=8.8cm]{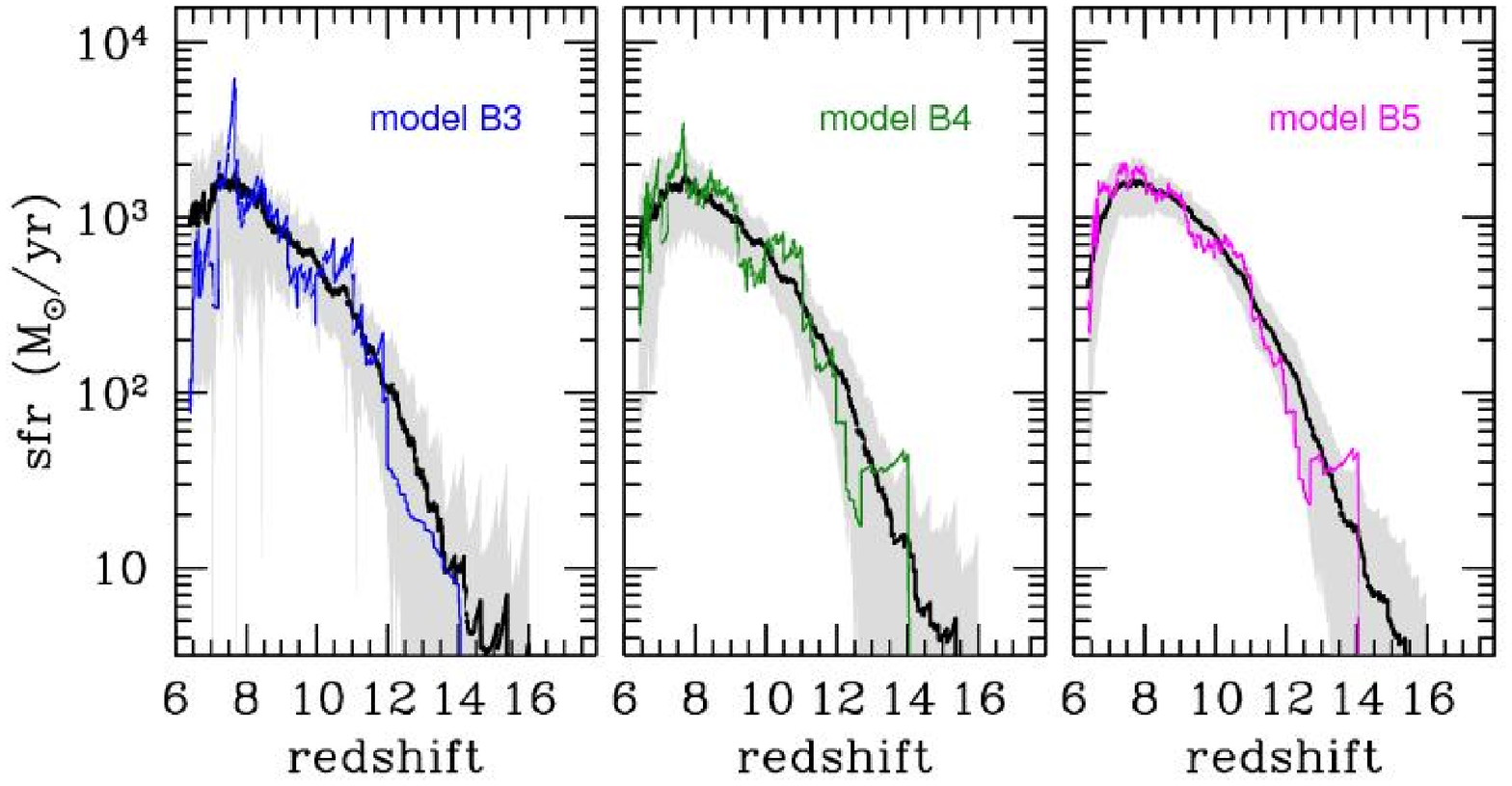}
  \caption{\small
    SFHs in high-$f_\ast$ models B3, B4 and B5 for a single merger tree 
    realization (thin lines). The panels show the effects of different values of
    the parameter $\sigma_{burst}$: 0.05 (B3, left panel), 0.1 (B4, central 
    panel) and 0.25 (B5, right panel). Thick solid lines refer to the average
    over 50 merger tree realizations (see text).} 
  \label{fig:sfrBsingle}
\end{figure}

With the aim of investigating the dependence of our results on the number and 
intensity of starbursts, we have explored two additional high-$f_\ast$ SFHs, 
models B4 and B5 (see Table \ref{tab:models2}).
These models predict a total stellar mass comparable to model B3 but with 
increasing values of $\sigma_{\rm burst}$ which imply SFHs characterized by an 
increasing number of (less efficient) bursts of star formation. 
The resulting average SFHs are shown in Fig.~\ref{fig:sfrBsingle} (thick lines
and corresponding 1-$\sigma$ dispersion).
The two SFH models have progressively less intense but more frequent bursts 
with respect to model B3.
These differences are particularly evident when we 
compare the predictions of a single merger tree realization; the thin lines 
in Fig.~\ref{fig:sfrBsingle} show the predicted SFHs in 
models B3, B4, and B5 for the same merger tree realization. Comparing with the 
corresponding thick lines in each panel, which represent the average SFH over
50 merger tree realizations, it becomes clear how the averaging
procedure partly cancel the redshift-dependent effects.  
\begin{table}
  \begin{center}
    \begin{tabular}{|c|c|c|c|c|c|}
      \hline
      {\bf model} & {\bf $\epsilon_{\rm quies}$} & {\bf $\sigma_{\rm burst}$} & {\bf $\epsilon_{\rm burst,max}$} & {\bf $\alpha$} & {\bf $\epsilon_{w,\textsc{agn}}$}\\
      \hline   \hline
      B4    & 0.1   &  0.1    & 4.0  &   233   &  $5\times10^{-3}$ \\
      B5    & 0.1   &  0.25   & 1.6  &   250   &  $5\times10^{-3}$ \\
      \hline
    \end{tabular}
  \end{center}
  \caption{\small 
    Model parameters of the two additional bursted SFH models, B4 and B5. 
    Column descriptions are the same as in Table \ref{tab:models}.}
  \label{tab:models2}
\end{table} 

\begin{figure}
  \centering
  \includegraphics[width=4.5cm]{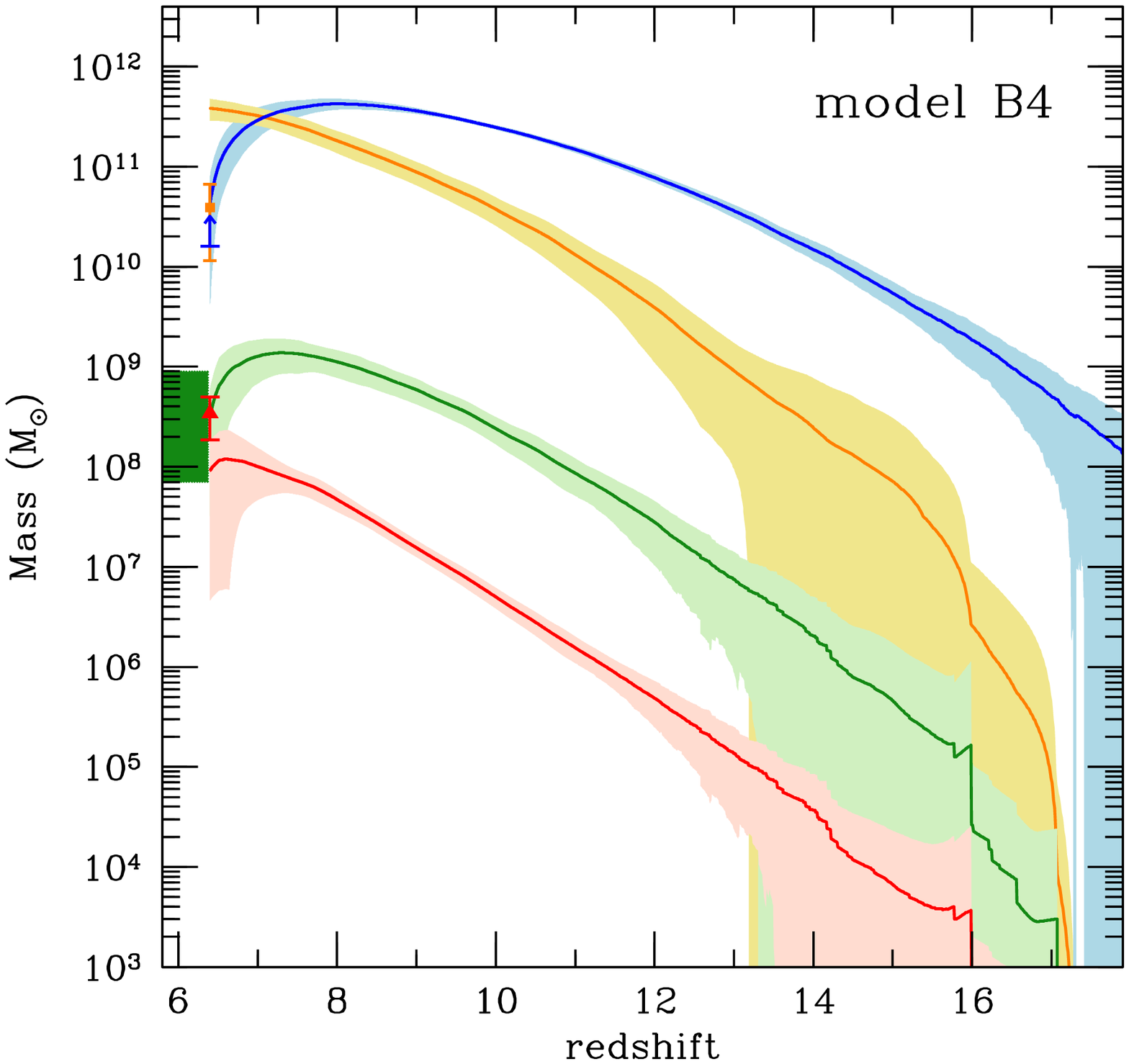}\includegraphics[width=4.5cm]{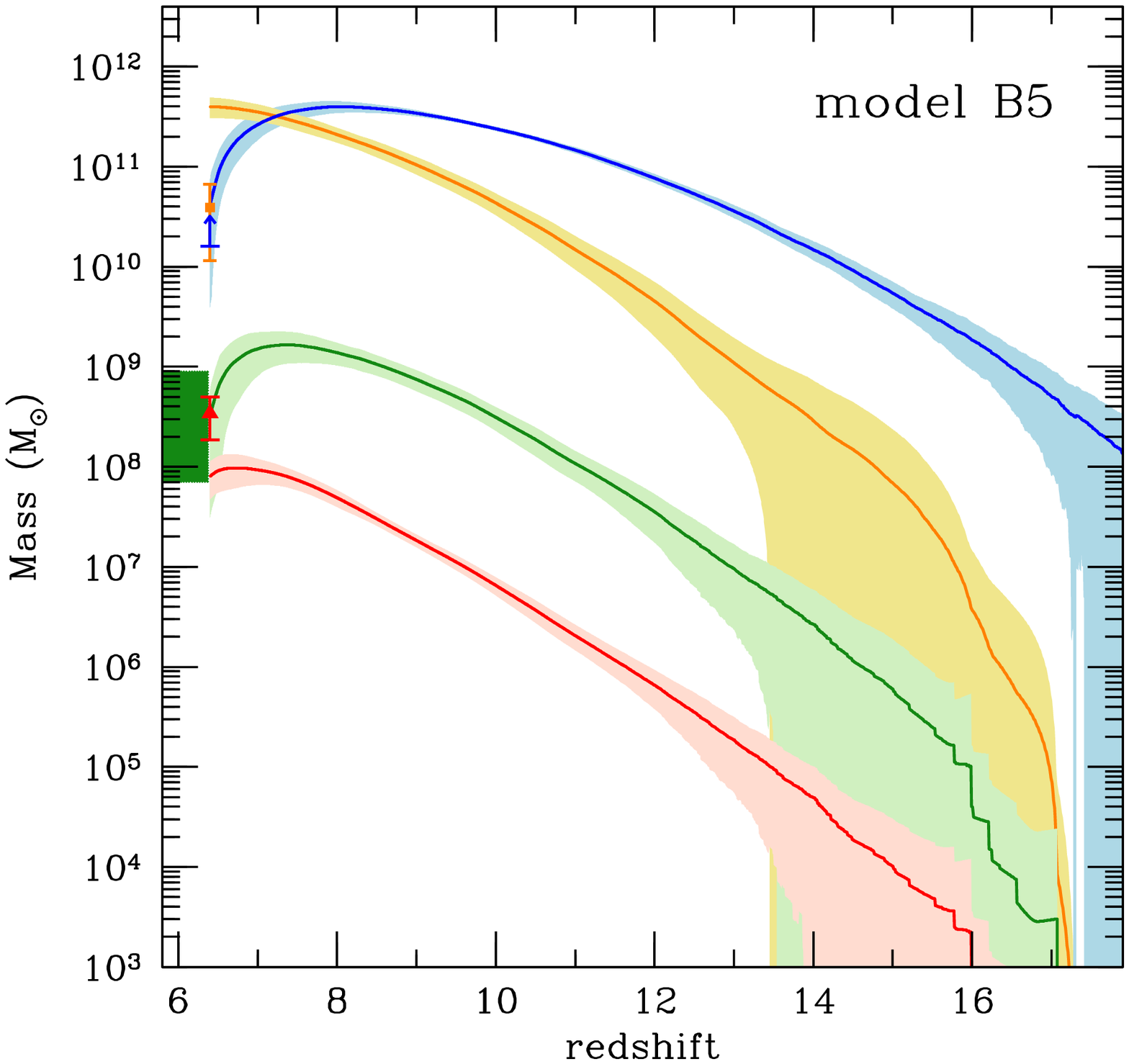}
  \caption{\small
    Evolution of the chemical properties (gas, 
    stars, metals and dust) for bursted SF models B4 and B5.
    Line colours and observational constraints are the same
    as in Fig.~\ref{fig:chemevo}.} 
  \label{fig:B4B5chemevo}
\end{figure}

The quasar host chemical evolution predicted by the additional models 
presented here, B4 and B5, are shown in Fig.~\ref{fig:B4B5chemevo}.
Even if the evolution at the highest redshifts is (slightly) different (the 
predicted masses are slightly higher and the curves are progressively 
smoother), the final masses of gas, stars and metals are very 
similar to what predicted by model B3 (right lower panel in 
Fig.~\ref{fig:chemevo}). 
On the other hand, we find that model B4 is in agreement with the observed 
dust mass within the $1\sigma$ dispersion on the average evolution (i.e. 
at least in one of the 50 merger histories the associated SFH is able to 
reproduce the observed properties of the quasar host), whereas 
model B5, which is characterized by a SFH with less intense and more frequent 
bursts, is not able to reproduce the observed dust mass.

\noindent
This analysis shows that: 
\begin{enumerate}
  \item By averaging the results over a set of merger tree realizations
    the specific features of each single SFH, such as the number, intensity
    and redshift distribution of starbursts and the intensity and position 
    of the burst triggered by the last major merger, are partly washed out. 
  \item An increasing number of starbursts does not produce a larger final 
    dust mass. The higher is the number of bursts (over the whole host 
    galaxy evolution), the smaller is their intensity, approaching a 
    "quiescent like\" \, SFH (see model B5); the final dust mass is strongly 
    limited by destruction (see section \ref{sec:dustresults}).
\end{enumerate}


\label{lastpage}

\end{document}